\documentstyle[12pt,axodraw]{article}
\begin{document}

\def\simgt{\mathrel{\lower2.5pt\vbox{\lineskip=0pt\baselineskip=0pt
           \hbox{$>$}\hbox{$\sim$}}}}
\def\simlt{\mathrel{\lower2.5pt\vbox{\lineskip=0pt\baselineskip=0pt
           \hbox{$<$}\hbox{$\sim$}}}}

\begin{titlepage}

\begin{flushright}
UCB-PTH-06-17 \\
LBNL-61484
\end{flushright}

\vskip 2.0cm

\begin{center}

{\Large \bf 
Supersymmetry without the Desert
}

\vskip 1.0cm

{\large Yasunori Nomura and David Poland}

\vskip 0.4cm

{\it Department of Physics, University of California,
                Berkeley, CA 94720} \\
{\it Theoretical Physics Group, Lawrence Berkeley National Laboratory,
                Berkeley, CA 94720} \\

\vskip 1.2cm

\abstract{Naturalness of electroweak symmetry breaking in weak scale 
supersymmetric theories may suggest the absence of the conventional 
supersymmetric desert.  We present a simple, realistic framework for 
supersymmetry in which (most of) the virtues of the supersymmetric 
desert are naturally reproduced without having a large energy interval 
above the weak scale.  The successful supersymmetric prediction 
for the low-energy gauge couplings is reproduced due to a gauged 
$R$ symmetry present in the effective theory at the weak scale. 
The observable sector superpotential naturally takes the form of 
the next-to-minimal supersymmetric standard model, but without 
being subject to the Landau pole constraints up to the conventional 
unification scale.  Supersymmetry breaking masses are generated by 
the $F$-term and $D$-term VEVs of singlet and $U(1)_R$ gauge fields, 
as well as by anomaly mediation, at a scale not far above the weak 
scale.  We study the resulting patten of supersymmetry breaking 
masses in detail, and find that it can be quite distinct.  We 
construct classes of explicit models within this framework, based 
on higher dimensional unified theories with TeV-sized extra dimensions. 
A similar model based on a non-$R$ symmetry is also presented. 
These models have a rich phenomenology at the TeV scale, and allow 
for detailed analyses of, e.g., electroweak symmetry breaking.}

\end{center}
\end{titlepage}

\section{Introduction}

Weak scale supersymmetry provides an elegant framework for explaining 
the origin of electroweak symmetry breaking.  In its simplest realization, 
one assumes that the fundamental scale of nature is extremely large, of 
order the Planck scale $M_{\rm Pl}$, and that supersymmetry is (dynamically) 
broken at a hierarchically small scale~\cite{Witten:1981nf}.  This picture 
of a supersymmetric desert, in fact, seems to be supported by the apparent 
unification of the three gauge couplings at a scale of $M_U \approx 
10^{16}~{\rm GeV}$.  Suppressions of various higher dimension operators, 
such as the ones leading to proton decay, are also naturally explained 
in this picture.

On the other hand, the fact that no definite sign of supersymmetry has 
been seen so far has put models based on the picture described above 
in a somewhat unpleasant situation.  Given the current experimental 
constraints, parameters of these models must typically be fine-tuned 
to reproduce the correct scale for electroweak symmetry breaking. 
Looking at this more carefully, the problem typically arises for 
(one of) the following reasons:
\begin{itemize}
\item
The Higgs mass squared parameter receives radiative corrections that 
are proportional to the top squark squared masses.  These corrections 
arise from the entire energy interval between the weak scale and 
the scale where the supersymmetry breaking masses are generated, and 
so are enhanced by the logarithm of the ratio of these two scales. 
For example, in the case where supersymmetry breaking is mediated by 
gravitationally suppressed interactions~\cite{Chamseddine:1982jx}, 
the logarithm is inevitably large, giving a large negative contribution 
to the Higgs mass squared parameter.  This leads to fine-tuning, since 
the large negative contribution must be canceled to a high degree 
by some positive contribution, such as the one coming from the 
supersymmetric mass term.
\item
The amount of cancellation required to reproduce the correct scale 
for electroweak symmetry breaking becomes smaller if the mass of the 
physical Higgs boson, $M_{\rm Higgs}$, becomes larger.  In the minimal 
supersymmetric standard model (MSSM), the value of $M_{\rm Higgs}$ 
is bounded from above by the $Z$ boson mass, $m_Z$, at tree level, 
so that we need an extra source of $M_{\rm Higgs}$ to make the 
Higgs boson significantly heavier than $m_Z$.  This is, however, 
not so easy to achieve, because the extra coupling (the Higgs quartic 
coupling) needed to make $M_{\rm Higgs}$ large is subject to the 
Landau pole constraint, and thus often not large enough to push 
up $M_{\rm Higgs}$ to the level enough to eliminate fine-tuning. 
\end{itemize}

It is, of course, possible to evade these difficulties and eliminate 
fine-tuning within the supersymmetric desert framework.  For example, 
the large logarithm between the weak and Planck scales may be avoided 
due to a special renormalization group property of moduli and anomaly 
mediated supersymmetry breaking, leading to a natural model of 
supersymmetry~\cite{Choi:2005hd,Kitano:2005wc}.  Alternatively, 
a large Higgs quartic coupling needed to obtain large $M_{\rm Higgs}$ 
may arise as a result of strong gauge dynamics, giving some of 
the low-energy states as composite particles~\cite{Harnik:2003rs,%
Chang:2004db,Birkedal:2004zx}.  Nevertheless, it is still true 
that none of these models are particularly simple.  The physics of 
electroweak symmetry breaking would be much simpler if all of the 
relevant physics occurred at energies not far from the weak scale.

In this paper, we thus take the viewpoint that the difficulties described 
above are suggestive hints for the absence of the supersymmetric desert. 
Note that these difficulties are associated with the existence of the 
supersymmetric desert, and not that of weak scale supersymmetry itself. 
Without the large desert, weak scale supersymmetry indeed allows for 
the possibility of constructing a fully natural model of electroweak 
symmetry breaking, adopting for example the scheme discussed recently 
in Ref.~\cite{Barbieri:2006bg}.  A natural question then is to what 
extent the successes of the conventional supersymmetric desert picture 
are preserved in such a scenario.  These include, in particular, a simple 
understanding of the weakness of gravity, the successful unification 
prediction for the low-energy gauge couplings, with the apparent unification 
scale close to the gravitational scale, and natural suppressions of 
certain higher dimension operators such as the ones leading to rapid 
proton decay and large neutrino masses.

In this paper we present a simple, realistic framework for supersymmetry 
in which (most of) the virtues of the supersymmetric desert are naturally 
reproduced without a large energy interval above the weak scale.  We show 
that the features usually attributed to the successes of the supersymmetric 
desert can in fact be preserved in a relatively simple setup with a low 
fundamental scale of $O(10\!\sim\!100~{\rm TeV})$.  Lowering the fundamental 
scale to the TeV region was first proposed in the context of solving the 
gauge hierarchy problem without supersymmetry~\cite{Arkani-Hamed:1998rs}. 
(For earlier work on lowering the fundamental scale, see~\cite{Witten:1996mz}.) 
The possibility of obtaining a prediction for the low-energy gauge couplings 
with a lowered fundamental scale was discussed in various (unification) 
scenarios~\cite{Dienes:1998vh,Bachas:1998kr,Ibanez:1999st}, and 
supersymmetry breaking with a TeV sized extra dimension was studied 
in~\cite{Antoniadis:1990ew,Pomarol:1998sd,Barbieri:2000vh}.  Mechanisms 
of suppressing unwanted operators were also considered, for example 
in~\cite{Arkani-Hamed:1999dc}.  We present in this paper a complete, 
effective field theory framework in which all of these issues are 
simultaneously addressed in a consistent manner.  The framework has 
several general implications, for example on the form of the observable 
sector superpotential and the pattern of supersymmetry breaking masses, 
leading to various interesting phenomenological consequences.  Classes 
of explicit models can be constructed within this framework.  We find 
it significant that fully realistic theories with a low fundamental 
scale are obtained, which allow for detailed phenomenological analyses, 
including electroweak symmetry breaking.

We construct our framework in two steps.  We first identify the structure 
of the theory at the weak scale.  This theory is supposed to be an effective 
field theory describing physics below the cutoff scale $M_c$.  The value 
of $M_c$ can be restricted by various phenomenological requirements; in 
particular, lower bounds of order a few TeV are obtained from the precision 
electroweak data in the context of particular models.  In this paper we 
mainly take $M_c$ to be between of order a few and a hundred TeV, to make 
the resulting theory the most natural in terms of electroweak symmetry 
breaking.  Higher values of $M_c$, however, are also possible.  We find that 
a $U(1)_R$ symmetry that assigns the same charge for all the matter and Higgs 
supermultiplets plays an important role in this framework.  In particular, 
we consider the gauging of this symmetry by canceling its mixed anomalies 
with the standard model gauge group by a nontrivial shift of a singlet 
(moduli) field.  We then find that the successful supersymmetric prediction 
for the low-energy gauge couplings is automatically reproduced, if the 
singlet field has a certain vacuum expectation value (VEV).  Generating 
the required VEV within the regime of effective field theory is nontrivial, 
but we demonstrate that it is possible.  An important consequence of this 
setup is that the superpotential for the Higgs fields takes the form of 
the next-to-minimal supersymmetric standard model: $W = \lambda S H_u H_d 
+ (\kappa/3) S^3$, and this is true even though the $U(1)_R$ symmetry is 
spontaneously broken at the cutoff scale.  (It is possible to consider 
an MSSM-type superpotential if there is additional dynamics generating 
the supersymmetric mass for the Higgs doublets.)  Since the fundamental 
scale of our theory is now of $O(10\!\sim\!100~{\rm TeV})$, however, 
the couplings $\lambda$ and $\kappa$ are not subject to the strong Landau 
pole constraint up to the conventional unification scale of $\approx 
10^{16}~{\rm GeV}$.  This is an interesting result.  Since the scale 
where the superparticle masses are generated is also very low (of order 
$M_c$), and so there is no large logarithmic running, the present framework 
provides a perfect platform for realizing the $\lambda$SUSY models of 
Ref.~\cite{Barbieri:2006bg}, discussed recently from the viewpoint of 
eliminating fine-tuning in electroweak symmetry breaking.  We discuss 
possible sources of supersymmetry breaking in our framework, and study 
their implications on the pattern of supersymmetry breaking masses.

We next seek possible theories above the scale $M_c$, which reduce to 
the effective theory described above below the energy scale of $M_c$. 
We construct classes of realistic models based on higher dimensional 
unified field theories, in which the standard model gauge fields propagate 
in an extra dimension(s) of size $\approx M_c^{-1}$.  These models provide 
an understanding of the universal contribution to the tree-level standard 
model gauge kinetic terms, which cannot be fully understood in the effective 
theory below $M_c$.  They also provide an explicit framework which allows 
us to relate supersymmetry breaking masses to the fundamental parameters 
of the theories.  We discuss the consistency of these models as effective 
field theories, including suppressions of proton decay and the absence 
of unphysical modes, and study the resulting pattern of supersymmetry 
breaking masses as well as the masses of the lightest Kaluza-Klein 
(KK) excitations.

The paper is organized as follows.  In the next section we describe the 
structure of the theory below $M_c$.  We show that the successful prediction 
for the gauge couplings is reproduced in this theory and that a consistent 
vacuum can be obtained.  In section~\ref{sec:pheno} we discuss phenomenological 
implications of the theory, especially those on supersymmetry breaking. 
In section~\ref{sec:model} we present classes of higher dimensional 
models describing physics above $M_c$ up to the fundamental scale $M_*$. 
We study their phenomenological implications, including the pattern 
of supersymmetry breaking masses and the spectrum of the lightest KK 
states.  Discussion and conclusions are given in section~\ref{sec:concl}. 
Appendix~\ref{app:soft} gives the expressions for soft supersymmetry breaking 
masses in the minimal higher dimensional model of section~\ref{sec:model}, 
and Appendix~\ref{app:non-R} presents an alternative class of theories 
in which a non-$R$ $U(1)$ gauge symmetry is used instead of the gauged 
$U(1)_R$ symmetry.

\section{Framework}
\label{sec:framework}

In this section we present our framework.  We first present a basic 
physical picture and then discuss some details about the viability 
of the framework.

\subsection{Basic picture}
\label{subsec:picture}

We consider that physics above a few hundred GeV is described by 
a four dimensional (4D) supersymmetric standard model.  The quadratic 
divergence for the Higgs mass squared parameter is then cut off, 
as usual, by loops of superparticles.  We assume that this theory 
is an effective field theory valid only below the cutoff scale 
$M_c = O(10\!\sim\!100~{\rm TeV})$, which is close (or equal) 
to the fundamental scale of nature, $M_*$.%
\footnote{The value of $M_c$ can be as low as a few TeV as long 
as it is consistent with the precision electroweak data, which is 
determined in the context of particular models above $M_c$.  We 
denote the range of $M_c$ between of order a few and a hundred TeV 
roughly as $M_c = O(10\!\sim\!100~{\rm TeV})$, throughout the paper.}
Our first question then is: what is the basic structure of this 
effective field theory?  In particular, we ask if there is a simple 
way of reproducing the successful supersymmetric prediction for 
the low-energy gauge couplings, and if so, what are the generic 
implications of it.

We consider, as usual, that the standard model quarks, leptons, and 
Higgs boson are embedded into chiral superfields: $Q_i$, $U_i$, $D_i$, 
$L_i$, $E_i$, $H_u$ and $H_d$, where $i=1,2,3$ is the generation index. 
The Yukawa couplings are then given by the superpotential:
\begin{equation}
  W_{\rm Yukawa} = (y_u)_{ij} Q_i U_j H_u + (y_d)_{ij} Q_i D_j H_d 
    + (y_e)_{ij} L_i E_j H_d.
\label{eq:Yukawa}
\end{equation}
Note that these are the only superpotential terms whose existence 
is (almost certainly) required from the low-energy data.  The 
``fundamental'' mass term for the Higgsinos, $W = \mu H_u H_d$, 
may or may not exist.  For example, if there is a singlet field $S$, 
the effective Higgsino mass term can arise from the superpotential 
term $W = \lambda S H_u H_d$ through the VEV of $S$, so that we do 
not need the term $W = \mu H_u H_d$.

The superpotential of Eq.~(\ref{eq:Yukawa}) possesses a $U(1)_R$ 
symmetry under which all the chiral superfields have a charge of 
$+2/3$, in the normalization that the superpotential carries a charge 
of $+2$.  This is, obviously, the unique $U(1)_R$ symmetry if we 
require that all the chiral superfields carry the same charge, which 
may have some motivation in the ultraviolet theory.  Now, suppose 
that we consider gauging this $U(1)_R$ symmetry (or its discrete 
subgroup $Z_{N,R}$ with sufficiently large $N$,%
\footnote{Given our normalization convention for the $R$ charges, 
$N$ of $Z_{N,R}$ does not have to be an integer.}
e.g. $N = 5$).  We then find that $U(1)_R$ (or $Z_{N,R}$) has the 
following mixed anomalies with respect to the standard model gauge 
group, $SU(3)_C \times SU(2)_L \times U(1)_Y$:
\begin{equation}
\begin{array}{llllll}
  U(1)_R\mbox{-}U(1)_Y^2: & A_1 & \equiv 
    & 3 \Bigl(-\frac{1}{3}\Bigr) \Bigl(\frac{3}{5}\Bigr)
        \Bigl(\frac{1}{6}+\frac{4}{3}+\frac{1}{3}+\frac{1}{2}+1\Bigr)
      + \Bigl(-\frac{1}{3}\Bigr) \Bigl(\frac{3}{5}\Bigr)
        \Bigl(\frac{1}{2}+\frac{1}{2}\Bigr)
    & = & -\frac{11}{5},
\\
  U(1)_R\mbox{-}SU(2)_L^2: & A_2 & \equiv 
    & 3 \Bigl(-\frac{1}{3}\Bigr) \Bigl(\frac{1}{2}\times 4\Bigr) 
      + \Bigl(-\frac{1}{3}\Bigr) \Bigl(\frac{1}{2}\times 2 \Bigr) + 2
    & = & -\frac{1}{3},
\\
  U(1)_R\mbox{-}SU(3)_C^2: & A_3 & \equiv 
    & 3 \Bigl(-\frac{1}{3}\Bigr) \Bigl(\frac{1}{2}\times 4\Bigr) + 3
    & = & 1,
\end{array}
\label{eq:U1R-anom}
\end{equation}
which must somehow be canceled.  (Note that the gauginos have a charge 
of $+1$ under $U(1)_R$, while the quarks, leptons and Higgsinos have 
a charge of $-1/3$.)  Here, we have taken the ``$SU(5)$-normalization'' 
for $U(1)_Y$, and have assumed that the MSSM quark, lepton and Higgs 
superfields are the only states charged under the standard model gauge 
group.  The $U(1)_Y$-$U(1)_R^2$ anomaly is automatically vanishing, 
and we do not consider the $U(1)_R^3$ or $U(1)_R$-$({\rm gravity})^2$ 
anomalies because they depend on unknown fields that are singlet under 
the standard model gauge group.

The anomalies of Eq.~(\ref{eq:U1R-anom}) can be canceled by the 
(generalized) Green-Schwarz mechanism~\cite{Green:1984sg}.  Assuming 
that a single moduli field $M$ is responsible for the cancellation, 
the interactions between $M$ and the standard model gauge fields 
are given by
\begin{equation}
  {\cal L} = -\sum_{I=1,2,3} \frac{A_I}{c} \int\!d^2\theta\, 
    M\, {\cal W}_I^{a\alpha} {\cal W}_{I\alpha}^a + {\rm h.c.},
\label{eq:gen-GS}
\end{equation}
where $I = 1,2,3$ represents $U(1)_Y$, $SU(2)_L$ and $SU(3)_C$, 
respectively, ${\cal W}_I^{a\alpha}$ the field strength superfields 
with $a$ representing the adjoint indices, and $c$ a real constant.%
\footnote{The field $M$ can be a linear combination of various moduli 
fields existing in the theory above $M_c$.}
The moduli field $M$ is normalized such that the coefficient of the 
$M^\dagger M$ term in the K\"ahler potential is of order the fundamental 
scale $M_*^2$, and transforms as $M \rightarrow M + i \alpha c /16\pi^2$ 
under $U(1)_R$, where $\alpha$ represents the $U(1)_R$ transformation 
parameter in the normalization that a chiral superfield with 
a charge $q$ transforms as $\Phi(x,\theta) \rightarrow e^{i q \alpha} 
\Phi(x,\theta e^{-i\alpha})$.  The coefficients in Eq.~(\ref{eq:gen-GS}) 
are then determined such that the anomalies of Eq.~(\ref{eq:U1R-anom}) 
are canceled by the classical transformation of Eq.~(\ref{eq:gen-GS}). 
(In the case of a discrete $Z_{N,R}$ symmetry, $\alpha = 2\pi/N$.) 
Here we choose the constant $c$ to be of $O(1)$.  This corresponds to 
giving a small ``charge'' of $O(1/16\pi^2)$ to $M$, and is necessary 
for the theory to stay within the regime of effective field theory 
(see the next subsection).  With the terms in Eq.~(\ref{eq:gen-GS}), 
we can gauge the $R$ symmetry, under which all the MSSM chiral 
superfields have a charge of $+2/3$.

We now consider the gauge couplings in this theory.  Since the 
coefficients of the terms in Eq.~(\ref{eq:gen-GS}) have indefinite signs, 
we need other positive gauge kinetic terms for (at least some of) the 
standard model gauge fields.  We assume that these terms are universal 
for the standard model gauge group (in the $SU(5)$ normalization), and 
that the gauge kinetic functions for $SU(3)_C$, $SU(2)_L$ and $U(1)_Y$ 
are given by the sum of the universal contribution and the ones from 
Eq.~(\ref{eq:gen-GS}).  Assuming that the universal piece arises from 
the VEV of a chiral superfield, which we denote by $T$, we obtain
\begin{equation}
  {\cal L} = \sum_{I=1,2,3} \int\!d^2\theta 
    \biggl( \frac{1}{4} T - \frac{A_I}{c} M \biggr) 
    {\cal W}_I^{a\alpha} {\cal W}_{I\alpha}^a + {\rm h.c.}
\label{eq:gauge-kin}
\end{equation}
This form requires some justification from the theory above the cutoff 
scale $M_c$; in particular, the normalization (coupling to $T$) of 
$U(1)_Y$ should be explained.  We will give examples of such theories 
in section~\ref{sec:model}.  Here we simply note that the form of 
Eq.~(\ref{eq:gauge-kin}) is technically natural.  As long as the VEV 
of $T$ is of order unity or larger (which is the case we are interested; 
see below), corrections to the gauge kinetic functions that do not 
respect the form of Eq.~(\ref{eq:gauge-kin}) are of order $1/8\pi^2$ 
or smaller, and thus are negligible for our purpose.

We assume that the VEVs of $T$ and $M$ are stabilized (dynamically) with 
$\langle T \rangle,\, \langle M \rangle = O(1)$.  (Note that the $T$ 
and $M$ fields are dimensionless in our convention.)  The standard model 
gauge couplings, $g_I$, at the scale $M_c = O(10\!\sim\!100~{\rm TeV})$ 
are then given by
\begin{equation}
  \frac{1}{g_I^2}(M_c) 
    = \langle T \rangle - \frac{4 A_I}{c} \langle M \rangle.
\label{eq:321-couplings}
\end{equation}
An important point here is that the anomaly coefficients $A_I$ are exactly 
proportional to the corresponding MSSM beta function coefficients $b_I$:
\begin{equation}
  \left( \begin{array}{c}
    b_1 \\ b_2 \\ b_3
  \end{array} \right)
  = \left( \begin{array}{c}
    33/5 \\ 1 \\ -3
  \end{array} \right)
  = -3 \left( \begin{array}{c}
    -11/5 \\ -1/3 \\ 1
  \end{array} \right)
  = -3 \left( \begin{array}{c}
    A_1 \\ A_2 \\ A_3
  \end{array} \right).
\label{eq:b-A-rel}
\end{equation}
We then find that the gauge couplings of Eq.~(\ref{eq:321-couplings}) 
satisfy exactly the same relation as that arising in the conventional 
supersymmetric desert picture:
\begin{equation}
  \frac{1}{g_I^2}(M_c) 
    = \frac{1}{g_U^2} + \frac{b_I}{8\pi^2} \ln\frac{M_U}{M_c},
\label{eq:321-GUT}
\end{equation}
where $g_U \simeq 0.7$ is the unified gauge coupling at the unification 
scale $M_U \simeq 2 \times 10^{16}~{\rm GeV}$.  (The gauge couplings 
in both Eqs.~(\ref{eq:321-couplings}) and (\ref{eq:321-GUT}) are the 
holomorphic gauge couplings.)  The explicit correspondence between 
the two theories is given by
\begin{eqnarray}
  \langle T \rangle 
    &\leftrightarrow& \frac{1}{g_U^2},
\label{eq:corresp-1}\\
  \langle M \rangle
    &\leftrightarrow& \frac{3 c}{32\pi^2}\ln\frac{M_U}{M_c},
\label{eq:corresp-2}
\end{eqnarray}
and the relation among the low-energy gauge couplings by
\begin{equation}
  \frac{1}{g_3^2} = \frac{12}{7}\frac{1}{g_2^2} - \frac{5}{7}\frac{1}{g_1^2}.
\label{eq:gcu}
\end{equation}
The relation of Eq.~(\ref{eq:gcu}) is, in fact, renormalization group 
invariant and well reproduces the observed QCD coupling, $g_3$, in 
terms of the electroweak gauge couplings, $g_1$ and $g_2$, at $m_Z$. 
The correspondence of Eq.~(\ref{eq:corresp-2}) implies that $M$ should 
be stabilized with $\langle M \rangle$ taking a positive value of $O(1)$. 
This is not entirely trivial to achieve and will be discussed in the 
next subsection.

In general, if we assign the $R$ charge of $+2/3$ for all the (charged) 
chiral superfields in the theory, we always obtain the relation $A_I 
= -b_I/3$ for any gauge group $I$ present in the theory.  Then, assuming 
that the mixed anomalies for the $R$ symmetry are canceled by the 
(generalized) Green-Schwarz mechanism with a single modulus $M$, 
as in Eq.~(\ref{eq:gauge-kin}), we always obtain the correspondence 
between the threshold effects from $\langle M \rangle$ and the running 
effect, given by Eq.~(\ref{eq:corresp-2}).  This originates from 
the relation between $R$ and dilatation symmetries in supersymmetric 
theories, although the $R$ symmetry considered here is not the exact 
supersymmetric partner of the (broken) dilatation symmetry.%
\footnote{Here we consider an $R$ symmetry that is not the exact 
supersymmetric partner of the (broken) dilatation symmetry but is 
a (unbroken) linear combination of it with some other $U(1)$ symmetry. 
The dependence of the superspace density, or the K\"ahler potential, 
on the $M$ field is then ``arbitrary'' in the effective field theory; 
in particular, the $U(1)_R$ gauge field $V_R$ and the combination 
$-(8\pi^2/c)(M+M^\dagger)$ can be used interchangeably as far as $U(1)_R$ 
gauge invariance is concerned.  Here we consider the case that the $M$ field 
appears simply in the gauge kinetic functions to cancel the mixed anomalies, 
so that its VEV leads to the large threshold effects without (much) affecting 
the superspace density.  In particular, we assume that $U(1)_R$ gauge 
invariance of the (observable sector) superspace density is recovered by 
the appropriate appearance of the $U(1)_R$ gauge field $V_R$, including 
the ``anomalous'' pieces.  (Note that the cutoff scale is ``charged'' under 
a part of the supersymmetric $U(1)_R$ gauge symmetry.)  This assumption 
should ultimately be understood in the ultraviolet theory above $M_*$ (or 
it perhaps suggests a certain structure for the ultraviolet theory), but 
it is a stable assumption in the framework of effective field theory.}
It is fortunate that the Yukawa couplings are dimensionless and thus 
allow this particular $R$ charge assignment.  (Other possible charge 
assignments, preserving the unification prediction, will be discussed 
briefly in section~\ref{sec:concl}.)

The prediction of Eq.~(\ref{eq:gcu}) receives corrections at higher 
orders if we take the $g_I$'s to be the canonically defined gauge 
couplings.  In particular, the prediction in our framework differs from 
that in the desert picture at higher orders, since some of the two-loop 
running effects between $M_U$ and $M_c$ are absent in our case.  (Part 
of the effects are retained through rescaling anomalies associated with 
the gauge multiplets~\cite{Arkani-Hamed:1997mj}.)  The difference, however, 
is small, of $O(1/8\pi^2)$, and is the same size as the effect arising 
from incalculable, nonuniversal corrections to Eq.~(\ref{eq:gauge-kin}).

Similar dynamics relating the low-energy gauge couplings to 
chiral anomalies were considered earlier in the context of string 
theory.  A pseudo-anomalous $U(1)$ gauge symmetry was considered in 
Ref.~\cite{Ibanez:1992fy} to obtain the weak mixing angle without 
grand unification, through the universal nature of the mixed gauge 
anomalies in weakly-coupled heterotic string theory.  A pseudo-anomalous 
gauge symmetry with nonuniversal mixed anomalies was considered in 
Ref.~\cite{Ibanez:1999st} in the context of (more general) string 
theory, in an attempt to lower the string scale (mainly) to an 
intermediate scale, although a proper implementation of the dynamics 
was not successfully realized.  Here we present a viable and realistic 
effective field theory framework, in which the fundamental scale can be 
lowered to the $(10\!\sim\!100)~{\rm TeV}$ region, preserving automatically 
the successful supersymmetric prediction for gauge coupling unification. 
As we have seen and will see in more detail in the next subsection, 
this provides nontrivial constraints on physics associated with $M$, 
e.g. the transformation property and the stabilization dynamics.

An important consequence of the present way of obtaining the prediction 
for the low-energy gauge couplings is that we cannot write a direct mass 
term for the Higgs doublets, $W = \mu H_u H_d$, since it is forbidden by 
the (gauged) $R$ symmetry.  Here we have assumed that the $M$ field does 
not appear in the tree-level superpotential, which may be justified in 
a theory above $M_c$.  The simplest way to generate the Higgsino mass 
term, then, is to introduce a singlet field $S$ which has a charge of 
$+2/3$ under the $R$ symmetry.  The most general superpotential is then 
given by 
\begin{equation}
  W = \lambda S H_u H_d + \frac{\kappa}{3} S^3 + W_{\rm Yukawa},
\label{eq:W}
\end{equation}
where $W_{\rm Yukawa}$ is given by Eq.~(\ref{eq:Yukawa}), and we have 
imposed the standard matter parity, or $R$ parity, under which $Q_i$, 
$U_i$, $D_i$, $L_i$ and $E_i$ are odd while $H_u$, $H_d$ and $S$ are 
even.  It is interesting that we are naturally led to the form of the 
superpotential of the next-to-minimal supersymmetric standard model. 
The Higgsino mass then arises from the VEV of $S$, which should be 
generated through supersymmetry breaking.  It is also interesting 
that higher dimension operators in the superpotential, such as the 
ones leading to rapid proton decay and large Majorana neutrino masses, 
are suppressed by the $R$ symmetry.%
\footnote{It will be necessary to have a constant term in the 
superpotential to cancel the cosmological constant after supersymmetry 
breaking, which can be regarded as a soft breaking term of the $R$ 
symmetry (arising dynamically).   This term can affect the form of 
the K\"ahler potential but not that of the superpotential, because 
of the supersymmetric nonrenormalization theorem, and our discussions 
are not affected by its existence.  The term, however, may affect 
the mass of the light pseudo-Goldstone boson state, which could arise 
from spontaneous $R$ symmetry breaking occurring associated with 
supersymmetry breaking.  The constant term in the superpotential 
will be discussed further in section~\ref{sec:pheno}. \label{ft:W0}}
(The possibility of generating mass terms, e.g. $W = \mu H_u H_d$, 
without using a singlet VEV will be discussed in the next subsection.)

Since the fundamental scale in our framework, $M_*$, is of order 
$10\!\sim\!100~{\rm TeV}$, the couplings $\lambda$ and $\kappa$ 
appearing in Eq.~(\ref{eq:W}) are not subject to the Landau pole 
constraint up to the unification scale $M_U$.  This allows us to have 
large couplings, e.g. $\lambda \simlt 2$ and $\kappa \simlt 1$, at 
the weak scale, which in turn allows us to have a large mass for the 
lightest Higgs boson, reducing fine-tuning.  In fact, it has recently 
been shown that a large value of $\lambda$ can eliminate fine-tuning 
in electroweak symmetry breaking while naturally preserving consistency 
with the precision electroweak data, because of extra contributions to 
the electroweak $T$ parameter coming from the Higgs-boson and Higgsino 
states~\cite{Barbieri:2006bg}.  For fine-tuning to be really eliminated, 
however, it is also necessary that there is no large logarithm between 
the weak scale and the scale where the superparticle masses are generated. 
Our framework also addresses this issue.  Since supersymmetry will be 
broken and mediated at the scale $\approx M_c$ (or $M_*$), there is 
no large logarithm between the mediation scale and the weak scale. 
The explicit pattern of supersymmetry breaking masses, and thus the 
form of the Higgs potential, depends on how supersymmetry is broken. 
In fact, there are many possible ways to incorporate supersymmetry 
breaking in our framework, and some of them will be discussed in the 
next section.  An explicit analysis of electroweak symmetry breaking 
in some of these supersymmetry breaking scenarios will be given 
in a separate publication~\cite{NP}.

We finally discuss physics associated with the $R$ symmetry and the 
moduli fields $T$ and $M$.  If the gauged $R$ symmetry is a continuous 
$U(1)_R$ symmetry, a nonvanishing and positive Fayet-Iliopoulos term of 
$O(M_{\rm Pl}^2)$ will be generated for $U(1)_R$~\cite{Chamseddine:1995gb}. 
Here, $M_{\rm Pl}$ appears because the observed gravitational scale is large, 
$M_{\rm Pl} \simeq 10^{18}~{\rm GeV}$, which may arise from the fact that 
gravity propagates in (large) spatial dimensions in which the MSSM states 
do not propagate~\cite{Arkani-Hamed:1998rs}.  We assume that this term 
is canceled by the VEV of some field $\phi$ that has a negative charge 
under $U(1)_R$.%
\footnote{Neither the MSSM scalars nor the $S$ scalar will obtain 
(disastrously large) VEVs from the $U(1)_R$ $D$-term, since they 
all have a $U(1)_R$ charge of $+2/3$.}
The kinetic term for this field must be enhanced by $M_{\rm Pl}^2/M_*^2$ 
so that the $\phi$ VEV does not far exceed the fundamental scale. 
Such an enhancement occurs if $\phi$ propagates in the same spacetime 
dimensions as the gravitational multiplet.  The $U(1)_R$ gauge supermultiplet 
then becomes massive, absorbing the $\phi$ supermultiplet.  The generated 
mass is only of order $M_*$, because the $U(1)_R$ gauge coupling must also 
receive a volume suppression of order $M_*/M_{\rm Pl}$ and the $U(1)_R$ 
gauge boson mass is given by the product of the $U(1)_R$ gauge coupling 
and the canonically normalized $\phi$ VEV.  (These issues will be elaborated 
on further in the next subsection.)  Depending on the $R$ charge 
of $\phi$, an unbroken discrete $R$ symmetry may remain at low energies.%
\footnote{An interesting alternative would be to cancel the Fayet-Iliopoulos 
term by the VEV of $M$ without introducing the $\phi$ field, which is 
a priori possible if the kinetic term of $M$ is enhanced by the factor 
of $M_{\rm Pl}^2/M_*^2$.  This is because the kinetic term of $M$ takes 
the form $\propto \int\!d^4\theta (M + M^\dagger + (c/8\pi^2)V_R)^2 + 
\cdots$, where $V_R$ is the $U(1)_R$ gauge supermultiplet, and thus gives 
a term linear in $M$ in the auxiliary component of $V_R$.  This, however, 
fixes the VEV of $M$ such that $c \langle {\rm Re}M \rangle < 0$, leading 
to a wrong prediction of $\ln(M_U/M_c) < 0$ in the correspondence of 
Eq.~(\ref{eq:corresp-2}).}

Supersymmetry breaking will also contribute to $U(1)_R$ breaking, 
because the gaugino masses violate $U(1)_R$, although by a much smaller 
amount than that from the $\phi$ (or $M$) VEV.  A potential $R$ axion from 
supersymmetry breaking will obtain a mass from operators involving the 
$\phi$ VEV, or from a (effective) constant term in the superpotential 
that should arise as a soft symmetry breaking term of $U(1)_R$ (see 
footnote~\ref{ft:W0}).  Masses for the $T$ and $M$ fields can be 
generated, i.e. their VEVs can be stabilized, through couplings of 
these fields to a gauge group(s) other than that in the standard model. 
This issue will be studied further in the next subsection. 

The story will be similar in the case that the gauged $R$ symmetry 
is a discrete $Z_{N,R}$ symmetry, except for the issues related to the 
$U(1)_R$ gauge multiplet and the $D$-term potential.  The $Z_{N,R}$ 
symmetry will be spontaneously broken through supersymmetry breaking 
(to the $Z_{2,R}$ subgroup), and the moduli fields $T$ and $M$ can 
still be stabilized by some gauge dynamics.

We note that all the VEVs and masses appearing in the analysis above 
can stay within the regime in which the effective field theoretic 
description is applicable (see the next subsection for more details). 
Despite the apparent appearance of the scale $M_{\rm Pl}$ ($\gg M_*$), 
no knowledge about physics above $M_*$ is required to describe the 
phenomena discussed above.

\subsection{Stabilizing moduli: producing the effective desert}
\label{subsec:M-stab}

As we have seen in the previous subsection, it is crucial that the 
$M$ field can be stabilized with $\langle M \rangle$ taking a positive 
value $\langle M \rangle/c \simeq 0.25$ (see Eq.~(\ref{eq:corresp-2})). 
The $T$ field should also be stabilized with $\langle T \rangle \simeq 2$ 
(see Eq.~(\ref{eq:corresp-1})).  In this subsection we discuss the issue 
of stabilization of these fields, concentrating on the case that the 
gauged $R$ symmetry is a continuous $U(1)_R$ symmetry.

For definiteness, we consider the case that the $M$ field does not 
propagate in large gravitational extra dimensions.  (The case in which 
it does can be treated similarly.)  In our convention, the $M$ field is 
dimensionless.  The K\"ahler potential for this field is then given by 
\begin{equation}
  K = M_*^2\, {\cal F}\biggl( M + M^\dagger + \frac{c}{8\pi^2}V_R \biggr),
\label{eq:K-M}
\end{equation}
where $V_R$ is the $U(1)_R$ gauge supermultiplet and ${\cal F}(x)$ 
is an arbitrary polynomial in $x$ with the coefficients of $O(1)$ 
up to symmetry factors.  The origin of $M$, $M=0$, is chosen such 
that the standard model gauge kinetic functions take the form of 
Eq.~(\ref{eq:gauge-kin}).  The form of Eq.~(\ref{eq:K-M}) immediately 
tells us that $c$ cannot be much larger than of order unity, since 
then the required value of $\langle M \rangle \simeq 0.25\, c$ would 
exceed $\simeq 1$, going outside the regime of effective field theory.

Let us first address the issue of the stabilization of $T$.  In general, 
the stabilization of $T$ is related dynamically to that of $M$, leading 
to a complicated potential minimization problem.  It is, however, possible 
that the two dynamics are practically decoupled.  The most straightforward 
way to realize that is to consider a gauge group(s) $G$ which does not 
have a mixed anomaly with $U(1)_R$.  This can be easily arranged by 
assigning appropriate $U(1)_R$ charges for the fields that transform 
nontrivially under $G$.  We can then use conventional mechanisms for 
dilaton stabilization to stabilize $T$.  For example, we can adopt one 
of the models discussed in Ref.~\cite{Izawa:1998dv}, which do not violate 
$U(1)_R$ invariance.  Alternatively, the stabilization of $M$ can be 
much stronger than that of $T$, in which case the stabilization of $T$ 
can be analyzed independently from that of $M$, after $M$ is fixed. 
We thus focus only on the stabilization of $M$ below, assuming that 
$T$ is independently stabilized.%
\footnote{An interesting, alternative possibility is to introduce 
extra vector-like matter states that are neutral under $U(1)_R$ 
and charged under the standard model gauge group.  These states 
then give extra contributions to the anomaly coefficients $A_I$ in 
Eq.~(\ref{eq:U1R-anom}).  Assuming that they have the quantum numbers 
of 2 pairs of ${\bf 5} + {\bf 5}^*$ of $SU(5) \supset SU(3)_C \times 
SU(2)_L \times U(1)_Y$, the extra contributions are $\delta\! A_I = -2$, 
leading to $(A_1, A_2, A_3) = (-21/5, -7/3, -1)$.  This can give 
the observed values of the standard model gauge couplings for 
$\langle M \rangle/c \simeq 0.25$, without ever introducing the 
$T {\cal W}_I^{a\alpha} {\cal W}_{I\alpha}^a$ term.  Masses for 
the extra states of order the weak scale or somewhat larger can 
be generated through K\"ahler potential terms (see discussions 
in section~\ref{sec:model}).  While this possibility is not realized 
in the explicit models of section~\ref{sec:model}, where $T$ always 
appears as a field parameterizing the size of an extra dimension(s), 
it may be a viable option if the theory just below $M_*$ is 4D 
(other than the gravitational dimensions).}

We consider the possibility that the $M$ field is stabilized strongly 
without using supersymmetry breaking effects.  This requires that the 
superpotential contains the effect of $U(1)_R$ breaking, since otherwise 
the $M$ dependence of the superpotential is completely fixed by $U(1)_R$, 
which does not allow the strong stabilization of $M$.  How does the 
effect of $U(1)_R$ breaking appear in the superpotential?  It can appear 
through the VEV of the field $\phi$ that absorbs the large Fayet-Iliopoulos 
term of $U(1)_R$: $\xi \simeq 2 M_{\rm Pl}^2$.  For the $\phi$ field to 
be able to absorb $\xi$, $\phi$ must propagate in large gravitational 
extra dimensions.  This is because if $\phi$ does not propagate in these 
dimensions, the $\phi$ VEV is simply bounded as $\langle \phi \rangle 
\simlt M_*$ in the effective field theory, so that it cannot absorb 
$\xi \gg M_*^2$.  On the other hand, if $\phi$ propagates in the large 
gravitational dimensions, its kinetic term is enhanced by the volume 
factor $M_{\rm Pl}^2/M_*^2$, so that the canonically normalized 4D 
(zero-mode) field $\hat{\phi} \equiv (M_{\rm Pl}/M_*)\phi$ can take a VEV 
as large as $M_{\rm Pl}$, and thus can absorb the large Fayet-Iliopoulos 
term of $O(M_{\rm Pl}^2)$.  Note that the $U(1)_R$ $D$-term potential 
is then given by
\begin{equation}
  V_D = \frac{g_R^2}{2} \frac{1}{\bigl( 1-|\hat{\phi}|^2/3M_{\rm Pl}^2 \bigr)^2}
    \left\{ \xi + \left( r_\phi-\frac{2}{3} \right)|\hat{\phi}|^2 \right\}^2,
\label{eq:VD-R}
\end{equation}
where $r_\phi$ ($< 0$) is the $U(1)_R$ charge of the $\phi$ 
field~\cite{Ferrara:1983dh}.%
\footnote{The $U(1)_R$ charge of the $\phi$ field, $r_\phi$, should 
be negative in order for the graviton kinetic term to have the 
correct sign at the vacuum.}
Here, we have assumed the minimal form of the superspace density for 
$\phi$, and $g_R$ is the 4D $U(1)_R$ gauge coupling, which receives 
a volume suppression of order $M_*/M_{\rm Pl}$.  The generated $U(1)_R$ 
gauge boson mass is of order $g_R \langle \hat{\phi} \rangle \simlt M_*$, 
which is compatible with the effective field theory treatment of 
the dynamics.

We now present an explicit example of a model stabilizing the $M$ field 
with the correct value of $\langle M \rangle$.  We consider a supersymmetric 
$SU(2)$ gauge theory with 4 ``quark'' chiral superfields ${\cal Q}_i$ 
($i=1,\cdots,4$).  We assume that the $U(1)_R$ charge of ${\cal Q}_i$'s 
is universal, which we denote by $r$.  The mixed $U(1)_R$-$SU(2)^2$ 
gauge anomaly coefficient is then given by $A = 4(1/2)(r-1)+2 = 2r$, 
and the gauge kinetic function for $SU(2)$ is given by
\begin{equation}
  {\cal L} = \int\!d^2\theta 
    \biggl( \frac{1}{4 g_0^2} - \frac{A}{c} M \biggr) 
    {\cal W}^{a\alpha} {\cal W}_\alpha^a + {\rm h.c.},
\label{eq:SU2-kin}
\end{equation}
where ${\cal W}^{a\alpha}$ is the field strength superfield for $SU(2)$ 
with $a$ representing the adjoint index.  The tree-level term $1/g_0^2$ 
may come from the VEV of some moduli field, which may or may not be $T$, 
and which we assume to be stabilized independently with $M$.  The values 
of $g_0$ can be naturally of $O(1)$, as for the standard model gauge group. 

There are six gauge-invariant meson operators constructed out of 
${\cal Q}_i$, which can be decomposed into a 5-plet, $({\cal Q} {\cal Q})_m$ 
($m=1,\cdots,5$), and a singlet, $({\cal Q} {\cal Q})$, under the 
$SP(4)$ subgroup of the flavor $SU(4)$ symmetry.  Nonperturbative $SU(2)$ 
dynamics induce VEVs for these operators $({\cal Q} {\cal Q})_m^2 + 
({\cal Q} {\cal Q})^2 = \Lambda^4$, where $\Lambda$ is the dynamical 
scale of $SU(2)$~\cite{Intriligator:1995au}.  We now introduce the 
superpotential term $W = k Z_m ({\cal Q} {\cal Q})_m$, where $Z_m$ 
is an $SU(2)$-singlet chiral superfield and $k$ a coupling constant. 
This leads to $\langle ({\cal Q} {\cal Q})_m \rangle = 0$ and $\langle 
({\cal Q} {\cal Q}) \rangle = \Lambda^2$, which can be used as a 
general scale generation mechanism through the $({\cal Q} {\cal Q})$ 
operator~\cite{Izawa:1997gs}.  For a sufficiently large value of $k$, 
this does not disturb possible other dynamics associated with 
$({\cal Q} {\cal Q})$.

We now use the dynamics described above to stabilize $M$.  We assume 
that the $\phi$ field, which absorbs the large Fayet-Iliopoulos term, 
has a $U(1)_R$ charge of $-2r$.  We then introduce the superpotential
\begin{equation}
  W = \eta X 
    \Biggl( \frac{\hat{\phi}}{M_{\rm Pl}} ({\cal Q} {\cal Q}) - M_X^2 \Biggr),
\label{eq:W-stab}
\end{equation}
where $X$ is a chiral superfield with a $U(1)_R$ charge of $+2$, $\hat{\phi}$ 
is the (canonically normalized) 4D mode of $\phi$, and $\eta$ and $M_X^2$ 
are coefficients of $O(1)$ and $O(M_*^2)$, respectively.  Here, the $X$ 
field, as well as the $SU(2)$ sector, are supposed not to propagate in 
the gravitational dimensions, and the $M_{\rm Pl}$ suppression in the 
first term in the bracket arises from the large volume factor associated 
with the $\phi$ field.  As discussed above, the $SU(2)$ dynamics 
effectively replaces $({\cal Q} {\cal Q})$ with the square of the 
dynamical scale $\Lambda$, which in turn is given by
\begin{equation}
  \Lambda = M_* e^{\frac{8\pi^2}{b g^2}}
  = M_* e^{-2\pi^2\left(\frac{1}{g_0^2}-8r\frac{M}{c}\right)},
\label{eq:Lambda}
\end{equation}
where $b = -4$ is the beta function coefficient for $SU(2)$, and $1/g^2 
= 1/g_0^2 - 4AM/c$ is the inverse-squared $SU(2)$ gauge coupling at the 
scale $M_*$.  We then obtain the supersymmetric minimum from the vanishing 
of the $D$- and $F$-term potential, given by Eqs.~(\ref{eq:VD-R},~%
\ref{eq:W-stab}).  In particular, the vanishing of $F_X = 
-(\partial W/\partial X)^*$ leads to $\exp\Bigl[-4\pi^2(1/g_0^2-8rM/c)\Bigr] 
= (M_X^2/M_*^2)(M_{\rm Pl}/\langle \hat{\phi} \rangle) = O(1)$, giving 
\begin{equation}
  \frac{\langle M \rangle}{c} = \frac{1}{8rg_0^2} 
    + O\biggl(\frac{1}{32\pi^2r}\biggr),
\label{eq:M-VEV}
\end{equation}
which stabilizes $M$.  The VEVs of the other fields are given by 
$\langle X \rangle = 0$ and $\langle \hat{\phi} \rangle = O(M_{\rm Pl})$, 
and the masses of the excitations around the minimum are all of order 
$M_*$, implying that the stabilization can be very strong.  The result 
obtained here, including the value of $\langle M \rangle$ given in 
Eq.~(\ref{eq:M-VEV}), is not affected if we replace the first term in 
the bracket of Eq.~(\ref{eq:W-stab}) by an arbitrary function of the 
$U(1)_R$-invariant combination $(\hat{\phi}/M_{\rm Pl})({\cal Q} {\cal Q})$. 

We find from Eq.~(\ref{eq:M-VEV}) that the phenomenologically required 
value of $\langle M \rangle/c \simeq 0.25$ can be obtained with a 
natural choice of parameters, $r g_0^2 \simeq 0.5$.  In particular, 
having $\langle M \rangle/c = O(1)$ is quite natural in the present 
stabilization mechanism.  This implies that the apparent closeness of 
the unification scale, $M_U$, and the gravitational scale, $M_{\rm Pl}$, 
can be naturally explained in the present context.  For example, if $g_0 
\simeq g_U$ (e.g. due to certain unification of $SU(2)$ with the standard 
model gauge group at or above $M_*$), then a natural choice of $r \simeq 1$ 
leads to the correct value for the apparent unification scale, $M_U 
\simeq 2 \times 10^{16}~{\rm GeV}$, in Eq.~(\ref{eq:corresp-2}).  In 
fact, the origin of this desired property is very simple and general. 
Let us imagine that the gauge coupling of the stabilizing gauge group 
$G$, as well as those of the standard model gauge group, are given 
by the sum of $O(1)$ contributions and the contributions from 
$\langle M \rangle$.  Then, if there is a superpotential interaction 
relating the dynamical scale of $G$ with some scale of order $M_*$ (as 
in Eq.~(\ref{eq:W-stab})), the value of $\langle M \rangle/c$ is fixed 
to be of order unity, which in turn implies that the apparent unification 
scale is hierarchically larger (or smaller) than the weak scale.  The 
gauge coupling of the stabilizing group $G$ is generically large at 
$M_*$, but can still stay within the field theory regime, for example 
by taking the relevant mass parameter in the superpotential ($M_X$ 
in Eq.~(\ref{eq:W-stab})) somewhat smaller than $M_*$.

We now discuss the robustness of our Higgs sector superpotential in 
Eq.~(\ref{eq:W}), and more generally the implication of $U(1)_R$ on 
the form of the observable sector superpotential.  Since $U(1)_R$ is 
broken strongly by the VEV of $\phi$, one might think that $U(1)_R$ 
invariance does not give any constraint on the form of the superpotential. 
However, since the superpotential is holomorphic in fields and the 
$U(1)_R$ charge of $\phi$ is negative, we find that no linear or 
quadratic term can appear in the observable sector superpotential 
through the VEV of $\phi$.  The form of the Higgs sector superpotential, 
Eq.~(\ref{eq:W}), is thus robust at the renormalizable level.  On 
the other hand, higher dimension operators can in general be induced 
through the $\phi$ VEV.  For example, if we choose $r=1$ in the example 
of the $M$ stabilization discussed above, the $U(1)_R$ charge of 
$\phi$ becomes $-2$, allowing e.g. the operator $W = k\, (L H_u)^2 H_u 
H_d/M_*^3$, which leads to Majorana masses for the observed neutrinos 
of $O(0.1~{\rm eV})$ for $k \sim 10^{-3}$ and $M_* \sim 100~{\rm TeV}$. 
(Possible proton decay operators, e.g. $W \sim QQQLH_u H_d$, should 
somehow be forbidden.)  The existence of higher dimension operators, 
however, is model dependent.  For example, if we choose the $U(1)_R$ 
charge of $\phi$ to be irrational, then no higher dimension operators 
are induced in the observable sector superpotential through the 
$\phi$ VEV.

The argument given above does not entirely exclude the possibility 
of having linear and/or quadratic terms in the observable sector 
superpotential.  For example, we can consider a supersymmetric $SU(2)$ 
gauge sector which has an identical structure to that used above in 
stabilizing $M$, but with the $U(1)_R$ charge of the ``quark'' fields 
${\cal Q}_i$ fixed to be $+1/3$.  This can lead to a mass term for 
the Higgs doublet $\mu \sim \Lambda^{\prime 2}/M_*$ through the tree-level 
superpotential coupling $W \sim ({\cal Q}{\cal Q})H_u H_d/M_*$, where 
$\Lambda'$ is the dynamical scale of $SU(2)$.  By choosing $\Lambda'$ 
and/or the coefficient of the superpotential operator appropriately, 
this allows us to reproduce the weak scale supersymmetric mass for 
the Higgs doublets without introducing a singlet field.  The dynamics 
described here, in fact, can also be applied in a theory with a singlet 
field $S$.  In this case, we obtain the Higgs sector superpotential 
$W_{\rm Higgs} = \lambda S H_u H_d + \mu H_u H_d + L_S^2 S + (M_S/2)S^2 
+ (\kappa/3) S^3$, where the second, third and forth terms arise from 
couplings to an $SU(2)$ gauge-invariant operator $({\cal Q}{\cal Q})$. 
An interesting property of this theory is that the mass parameters 
appearing in the superpotential are naturally of the same order, 
$\mu \sim L_S \sim M_S \sim \Lambda^{\prime 2}/M_*$, which can be 
taken to be of order the weak scale by appropriately choosing the 
value of $\Lambda'$.  This, therefore, can be used to realize a general 
$\lambda$SUSY setup discussed in Ref.~\cite{Barbieri:2006bg}.  In the 
rest of the paper, however, we focus for simplicity on the case of 
Eq.~(\ref{eq:W}), which does not require additional dynamics generating 
dimensionful parameters in the observable sector superpotential.

\section{Phenomenological Implications}
\label{sec:pheno}

In this section we discuss general phenomenological implications of the 
framework described in the previous section.

\subsection{Supersymmetry breaking}
\label{subsec:susy-br}

There are in general many possible ways to incorporate supersymmetry 
breaking in our framework.  Here we identify several sources of 
supersymmetry breaking, intrinsic to our setup.  We study the resulting 
superparticle spectrum and its phenomenological implications.

In general, any field that is singlet under the standard model gauge 
group has the potential to provide supersymmetry breaking effects in the 
observable sector, through its auxiliary component VEV.  In our context, 
natural candidates are given by the VEVs of the auxiliary components of 
the $T$ and $M$ supermultiplets, $F_T$ and $F_M$.  Nonvanishing values 
for $F_T$ and $F_M$ can be generated through the stabilization mechanisms 
of $T$ and $M$.  Here we study their phenomenological implications 
without specifying explicit dynamics generating these VEVs.  For earlier 
related studies, see e.g.~\cite{Ibanez:1992hc}.

The couplings of the $T$ and $M$ superfields to the $SU(3)_C \times SU(2)_L 
\times U(1)_Y$ gauge multiplets are given by Eq.~(\ref{eq:gauge-kin}). 
This gives a definite prediction for the gaugino masses.  At the scale 
$M_c$, the $SU(3)_C \times SU(2)_L \times U(1)_Y$ gaugino masses, 
$M_I$ ($I=1,2,3$), are given by
\begin{equation}
  \frac{M_I}{g_I^2} = -\frac{1}{2}F_T - \frac{2b_I}{3c}F_M,
\label{eq:M_I}
\end{equation}
where $g_I$ are the standard model gauge couplings at $M_c$, and we have 
used Eq.~(\ref{eq:b-A-rel}).  In fact, since $M_I/g_I^2$ are renormalization 
group invariants, the gaugino masses at an arbitrary scale $\mu_R$ are 
given by Eq.~(\ref{eq:M_I}) with $g_I$ interpreted as the standard model 
gauge couplings at the scale $\mu_R$.  In the limit $F_M \rightarrow 0$, 
these gaugino masses reproduce the ones arising from the standard ``unified 
gaugino mass assumption'': $M_I \propto g_I^2$.  In the other extreme 
limit of $F_T \rightarrow 0$, the gaugino masses satisfy $M_I \propto 
b_I g_I^2$, the same relation as that in anomaly mediated supersymmetry 
breaking~\cite{Randall:1998uk}.  The effects of real anomaly mediation 
in the present context will be discussed later.

Since the three gaugino masses, $M_I$, are determined by two free parameters 
$F_T$ and $F_M/c$, we have one relation among them:
\begin{equation}
  \frac{M_3}{g_3^2} = \frac{12}{7}\frac{M_2}{g_2^2} 
    - \frac{5}{7}\frac{M_1}{g_1^2},
\label{eq:MI-rel}
\end{equation}
regardless of the values of $F_T$ and $F_M/c$, where we have used 
$(b_1, b_2, b_3) = (33/5, 1, -3)$.  The ratios between two of the $M_I$'s, 
e.g. $M_2$ and $M_3$, depend on the ratio between $F_T$ and $F_M/c$. 
An interesting property for the gaugino masses in Eq.~(\ref{eq:M_I}) (or 
Eq.~(\ref{eq:MI-rel})) which appears if $F_T$ and $F_M$ are real (more 
precisely, if the complex phases of $F_T$ and $F_M$ are the same) is that 
when these masses, as well as the gauge couplings, are extrapolated naively 
to high or low energies using the MSSM renormalization group equations, 
the three gaugino masses $M_I$ (not $M_I/g_I^2$) meet at a point at some 
(fictitious) energy scale.  The scale where they meet depends on the 
ratio between $F_T$ and $F_M/c$, and is given by
\begin{equation}
  M^\lambda_U = M_U\, 
    \exp\left(-\frac{8\pi}{3\alpha_U}\frac{F_M}{cF_T}\right),
\label{eq:gaugino-unif}
\end{equation}
where $M_U$ is the conventional gauge coupling unification scale, 
$M_U \simeq 2 \times 10^{16}~{\rm GeV}$, and $\alpha_U \equiv 
g_U^2/4\pi$ the unified gauge coupling strength, $\alpha_U \simeq 
1/24$.  From a purely low-energy point of view, this phenomenon is 
reminiscent of that in mixed moduli-anomaly mediated supersymmetry 
breaking~\cite{Choi:2004sx}, although the underlying physics picture 
is very different.  For $F_T F_M > 0$ ($< 0$), the {\it effective 
gaugino mass unification scale}, $M^\lambda_U$, is below (above) the 
effective gauge coupling unification scale $M_U$.  In particular, for 
$F_T/(F_M/c) \simgt 6$, $M^\lambda_U$ is between $M_U$ and the weak 
scale $\approx 100~{\rm GeV}$.  Note, however, that from a theoretical 
point of view there is no particular reason that $M^\lambda_U$ must 
be in this region.

The couplings of the $T$ and $M$ superfields to the matter and Higgs 
superfields are not determined within the effective field theory below 
$M_c$.  In section~\ref{sec:model} we give models in which $T$ is 
identified as the radion superfield associated with an extra dimension(s) 
in which the standard model gauge fields propagate.  The couplings 
of $T$ to the matter and Higgs fields then depend on the wavefunction 
profiles of these fields in the extra dimension(s), as well as the 
higher dimensional spacetime curvature.  The couplings of $M$ can 
also contain arbitrary functions of $M + M^\dagger + (c/8\pi^2)V_R$ 
in the effective theory.  The issue of supersymmetric flavor changing 
neutral currents should thus be addressed in a theory at or above $M_c$. 
The models in section~\ref{sec:model} provide examples of such a 
framework.  An alternative possibility is to consider some flavor 
symmetry, ensuring flavor universality for the squark and slepton 
masses. 

Another interesting source of supersymmetry breaking in our framework 
comes from a possible non-decoupling $U(1)_R$ $D$-term.  (For earlier 
work on supersymmetry breaking in a theory with gauged $U(1)_R$, see 
e.g.~\cite{Kitazawa:2000ft}.)  Suppose, for example, that the field 
canceling the $U(1)_R$ Fayet-Iliopoulos term, $\phi$, has a supersymmetry 
breaking mass squared $m^2$ of order the weak scale or somewhat larger 
(in the basis where $\phi$ is canonically normalized).  Such a mass can 
arise from the VEV of $F_T$ (and/or $F_M$) if the $T$ (and/or $M$) field 
propagates in the large gravitational dimensions, in which case the size 
of $m$ can naturally be of the same order as other supersymmetry breaking 
masses arising from $F_T$ (and/or $F_M$).  In this case, the minimization 
of the potential leads to a nonvanishing $D$-term VEV for $U(1)_R$, 
$D_R = O(m^2)$, {\it regardless of the value of the $U(1)_R$ gauge 
coupling}.  This gives a supersymmetry breaking squared mass of 
$\approx (r_i - 2/3)(-D_R)$ to a scalar field that has a $U(1)_R$ charge 
of $r_i$ through the $U(1)_R$ $D$-term potential ($-D_R$ is positive in 
our notation).  Since all the quark, lepton and Higgs superfields have 
a $U(1)_R$ charge of $+2/3$, however, this contribution is absent in 
our theory.%
\footnote{The theory allows us to write a kinetic mixing term between the 
$U(1)_R$ and $U(1)_Y$ gauge fields at tree level.  The supersymmetry breaking 
squared masses for the scalars can then obtain contributions proportional 
to their $U(1)_Y$ hypercharges.  If the mixing term has an $O(1)$ coefficient 
in the basis where the gauge couplings appear in front of the kinetic terms, 
these contributions can be of order $m^2$.  Note that a coefficient of 
$O(1)$ is phenomenologically harmless, since this term is suppressed 
by $g_R = O(M_*/M_{\rm Pl})$ in the basis where the gauge fields are 
canonically normalized.}
A nonvanishing contribution may arise if there are direct couplings 
of the form $\int\!d^4\theta (\phi^\dagger e^{2r_\phi V_R} \phi) 
(Q_i^\dagger e^{4V_R/3} Q_j)$ in the superspace density, where $Q_i$ 
represents generic quark, lepton and Higgs chiral superfields.  Since 
these couplings are not flavor universal in general, we may need to impose 
a nontrivial flavor symmetry if they give nonnegligible contributions. 

A nonvanishing $U(1)_R$ $D$-term VEV also gives a contribution of 
$(2/3)\gamma_i D_R$ to the scalar squared masses, since the cutoff 
is ``charged'' under a part of the supersymmetric $U(1)_R$ gauge 
symmetry (i.e. we must include terms involving $V_R$ to cancel 
``anomalous'' variations of the superspace density under $U(1)_R$ 
transformations).  Here, $\gamma_i$ represents the anomalous dimension 
of $Q_i$, defined by $d\ln Z_i/d\ln\mu_R \equiv -2\gamma_i$ with 
$Z_i$ the wavefunction renormalization for $Q_i$.  This gives a positive 
and approximately flavor universal contribution to the first two 
generation squark and slepton squared masses (at the scale $\approx 
M_c$), which becomes important if the value of $\sqrt{|D_R|}$ is somewhat 
larger than the weak scale and if the direct couplings between $\phi$ 
and $Q_i$ are suppressed in the superspace density, e.g. by locality 
in an extra dimension.%
\footnote{In fact, this contribution can naturally be the dominant 
one if we do not introduce $\phi$ from the beginning, since then 
$D_R = O(M_*^2)$.  In that case, however, we must come up with 
an alternative model for the $M$ stabilization.}
While this contribution leads to some amount of flavor violation, 
especially in the top squark sector, it is sufficiently small. 
Note that the contributions from a nonvanishing $U(1)_R$ $D$-term 
VEV discussed above preserve the gaugino mass prediction of 
Eq.~(\ref{eq:MI-rel}). 

The contribution from anomaly mediation~\cite{Randall:1998uk} can 
also be sizable if the $T$ (and/or $M$) field propagates in the large 
gravitational dimensions.  Suppose that (one of) the dominant contribution 
to observable sector supersymmetry breaking comes from $F_T$, and that 
$T$ propagates in all the gravitational dimensions.  This is indeed 
the case if $T$ parameterizes the size of an extra dimension(s), as in 
the models discussed in the next section.  Let us now consider the 4D 
effective theory obtained after integrating out all the extra dimensions. 
In this theory, the positive contribution to the vacuum energy from 
supersymmetry breaking $\delta V$ is of order $F_T^2 M_{\rm Pl}^2$, 
which is canceled by a (effective) constant term in the superpotential 
$\langle W \rangle = O(F_T M_{\rm Pl}^2)$, where $F_T$ is of order the 
weak scale.  (Both $F_T$ and $\langle W \rangle$ should arise from dynamical 
breaking of the $R$ symmetry.)  Note that these values of $\delta V$ and 
$\langle W \rangle$ ($\delta V \gg M_*^4$ and $\langle W \rangle \gg M_*^3$) 
are consistent with the effective field theory treatment of the dynamics, 
as the apparent large scales arise simply from the large volume factor 
associated with the large gravitational dimensions.  This implies that 
the effective bulk cosmological constant is negative before supersymmetry 
breaking.  The $F$-term VEV of a chiral compensator field $F_C$, which 
controls the size of anomaly mediation, depends on the mechanism of 
$T$ stabilization, but it typically takes a value in the range $F_T 
\simlt F_C \simlt 8\pi^2 F_T$.  (For analyses in the context of the 
conventional desert framework, see e.g.~\cite{Luty:1999cz}.)  Since 
the anomaly mediated contribution to the observable sector superparticle 
masses, $m_{\rm AMSB}$, is of order $F_C/8\pi^2$, we obtain
\begin{equation}
  \frac{F_T}{8\pi^2} \simlt m_{\rm AMSB} \simlt F_T.
\label{eq:amsb}
\end{equation}
We thus find that the contribution from anomaly mediation can be 
comparable to that from $F_T$ (and $\sqrt{|D_R|}$) in the present 
framework.%
\footnote{It is possible that the dynamics stabilizing $T$ is localized 
to a subspace in the direction of the gravitational dimensions.  In 
this case the mass of the $T$ field is of order $M_c^2/M_{\rm Pl} = 
O(0.01\!\sim\!10~{\rm eV})$, because of the volume suppression factor 
associated with the gravitational dimensions.  The wavefunction of this 
state can be (highly) nontrivial if the mass is close to the scale of 
the gravitational dimensions $L^{-1}$, which can occur if the number 
of these (flat) dimensions is two.  On the other hand, the VEVs of $T$ 
and $F_T$ always tend to have constant profiles along these dimensions 
(in the flat space case), because of the associated kinetic energy. 
Therefore, the low-energy, or zero-energy, 4D consideration leading 
to Eq.~(\ref{eq:amsb}) is still valid in this case.}
An interesting point is that since the anomaly mediated contribution 
satisfies $M_I \propto b_I g_I^2$, it does not destroy the gaugino mass 
prediction of Eq.~(\ref{eq:MI-rel}).  ($F_M/c$ in Eq.~(\ref{eq:gaugino-unif}) 
should be replaced by $F_M/c + 3F_C/32\pi^2$.)  The gravitino mass 
is given by $m_{3/2} \simeq F_C$, which is typically in the range of 
$O(100~{\rm GeV}\!\sim\!10~{\rm TeV})$.  Note that if the dominant 
contribution to observable sector supersymmetry breaking comes from 
a field that does not propagate in the gravitational dimensions, the 
gravitino mass is very small $m_{3/2} = O(m_{\rm weak}^2/M_{\rm Pl})$, 
and the anomaly mediated contribution, $m_{\rm AMSB} \simeq 
m_{3/2}/8\pi^2$, is completely negligible.

We finally comment on other possibilities for supersymmetry breaking in 
our framework.  We can consider that the dominant source of supersymmetry 
breaking comes from the auxiliary field VEV(s) of a chiral superfield(s) 
other than $T$ and $M$.  For example, we can consider a chiral superfield 
$Z$ which couples to the standard model gauge fields as $\int\!d^2\theta\, 
Z\, {\cal W}_I^{a\alpha} {\cal W}_{I\alpha}^a$ with arbitrary coefficients 
for $SU(3)_C$, $SU(2)_L$ and $U(1)_Y$.  (The lowest component VEV of 
$Z$ should be small/vanishing in order not to contribute to the gauge 
couplings.)  This can give arbitrary masses for the gauginos, which 
do not respect Eq.~(\ref{eq:MI-rel}).  In fact, this scenario can be 
naturally accommodated in a higher dimensional scenario discussed in 
section~\ref{subsec:higher-d}.  The couplings of $Z$ to the matter and 
Higgs fields can be naturally suppressed, leading to the spectrum of 
gaugino mediation~\cite{Kaplan:1999ac} but with a very low compactification 
scale $M_c = O(10\!\sim\!100~{\rm TeV})$.  The Higgs fields and (a part 
of) the third generation scalars may have different masses than the other 
scalars.  An interesting property of this model is that the gauginos are 
significantly heavier than the scalar particles, typically by a factor 
of a few.  We leave detailed studies of these and related possibilities 
for future work.

\subsection{Gravity, proton decay, and neutrino masses}
\label{subsec:others}

Since the fundamental scale of nature, $M_*$, is of order 
$M_c \approx (10\!\sim\!100)~{\rm TeV}$ in our framework, suppressions 
of various operators and interactions must be explained without 
using energy scales larger than $M_*$.  Here we list several 
possibilities for achieving this.

The weakness of gravity can be explained if there are large 
gravitational dimensions in which the MSSM or $S$ states do not 
propagate~\cite{Arkani-Hamed:1998rs}.  Assuming that the sizes of 
these dimensions are (approximately) equal, we find $L^{-1} \approx 
(M_*^{2+n}/M_{\rm Pl}^2)^{1/n}$, giving $L^{-1} \approx 10^{-11} - 
1~{\rm GeV}$ ($L \approx 10^{-16} - 10^{-5}~{\rm m}$) for $n = 
2,\cdots,6$ and $M_* = (10\!\sim\!100)~{\rm TeV}$.  Here, $n$ is 
the number of extra gravitational dimensions, and we have assumed 
that the extra space is flat.  (In the case that supersymmetry is 
broken by the auxiliary field VEV of a bulk supermultiplet, such 
as a radion, this implies that the bulk cosmological constant is 
negative in the limit of unbroken supersymmetry, which is canceled 
by a positive contribution from bulk supersymmetry breaking.) 
Dimensions of these sizes are not constrained by the existing 
submillimeter gravitational experiments, although there are 
astrophysical constraints for $n = 2$~\cite{Cullen:1999hc}.

Proton decay should be suppressed much more strongly than what is 
naively expected based on the scale $M_*$.  In section~\ref{sec:model}, 
we present models above $M_c$ based on higher dimensional unified 
field theories.  In these models there exist KK states for the unified 
gauge fields and colored Higgs multiplets, and the requirement that 
proton decay should not be caused by the exchange of these states gives 
nontrivial constraints on the structure of the models.  There are also 
possible tree-level proton decay operators.  These can be suppressed 
if the quark and lepton supermultiplets are localized at different 
positions in extra dimensions.  The relevant dimension for the 
separation can be one of the large gravitational dimensions as 
originally considered in~\cite{Arkani-Hamed:1999dc}, but can also 
be a dimension of $O(M_c^{-1})$, orthogonal to the ones used in 
explaining the weakness of gravity.  A model accommodating such 
a possibility will be presented in section~\ref{subsec:higher-d}. 
Alternative ways of suppressing these operators include imposing 
an appropriate (gauged) discrete symmetry, or a continuous gauged 
baryon and/or lepton number broken on a distant brane.

Small neutrino masses can be generated by introducing right-handed 
neutrino superfields $N$ having an $R$ charge of $+2/3$.  Note that 
dangerous superpotential operators $W \sim (L H_u)^2$, giving too 
large Majorana neutrino masses, are forbidden by the $R$ symmetry, and 
they are not regenerated unless there is a nonvanishing VEV carrying 
an $R$ charge of $-2/3$ (assuming that the $M$ field does not appear 
in the superpotential).  If the $N$ fields propagate in (a part of) 
the large gravitational dimensions, the 4D neutrino Yukawa couplings 
$W \sim L N H_u$ are suppressed by a factor of $(M_*/M_{\rm Pl})^{m/n}$, 
where $m$ is the number of dimensions in which $N$ propagates, giving 
naturally small Dirac neutrino masses~\cite{Arkani-Hamed:1998vp}. 
For example, a neutrino mass of $O(0.01\!\sim\!0.1~{\rm eV})$ relevant 
for atmospheric neutrino oscillations is naturally obtained for $m=n$ 
and $M_* \simeq 100~{\rm TeV}$.  Alternatively, small Majorana neutrino 
masses may be generated from higher dimension operators, such as 
$W \sim (L H_u)^2 H_u H_d$, through the $\phi$ VEV, as discussed 
in section~\ref{subsec:M-stab}.

In general, large mass scales can be obtained effectively for any 
higher dimension operators by making the relevant field(s) propagate 
in large extra dimensions.  For example, the QCD axion can be obtained 
by coupling the axion superfield $\Phi$ to the $SU(3)_C$ gauge field as 
$\int\!d^2\theta\, \Phi\, {\cal W}_3^{a\alpha} {\cal W}_{3\alpha}^a$, 
and making $\Phi$ propagate in (a part of) the large gravitational 
bulk~\cite{Arkani-Hamed:1998nn}.  The (effective) axion decay constant 
is then given by $f_a \approx M_* (M_{\rm Pl}/M_*)^{m/n}$, where $m$ 
is the number of dimensions in which $\Phi$ propagates.

With $M_* = O(10\!\sim\!100~{\rm TeV})$, most other higher dimension 
operators are phenomenologically harmless.  The operators leading 
to flavor changing neutral currents, however, also need some 
suppressions.  For instance, the coefficients of the operators 
leading to the $K^0$-$\bar{K}^0$ mixing must be smaller than of order 
$10^{-2}(M_*/100~{\rm TeV})^2$ in units of $M_*$.  (The coefficients 
must be even smaller by a factor of $\approx 100$ if they have $O(1)$ 
phases.)  The origin of these suppressions is presumably related to 
the physics giving the Yukawa couplings.  For example, we can consider 
the situation in which the wavefunction renormalization factors for 
lighter generation quarks and leptons are enhanced compared to the 
heavier ones.  After canonically normalizing the fields, this leads 
to realistic Yukawa couplings as well as suppressions of flavor 
changing higher dimension operators.  In fact, this situation can 
easily be realized if lighter generation quarks and leptons propagate 
in extra dimensions somewhat larger than $M_*^{-1}$.  An alternative 
possibility to suppress flavor changing neutral currents is to impose 
a flavor symmetry whose breaking resides only in the Yukawa couplings.

\section{Explicit Models}
\label{sec:model}

In this section we present possible theories above $M_c$ which 
reproduce the effective theory below $M_c$ discussed in the previous 
sections.  In particular, we present models in which the structure 
of Eq.~(\ref{eq:gauge-kin}), especially the universal coupling of the 
standard model gauge fields to $T$, is naturally reproduced.  Below 
we provide models based on 5D spacetime, in which the standard model 
gauge fields propagate.  The existence of additional (orthogonal) 
large gravitational dimensions, however, should be understood as 
discussed in section~\ref{subsec:others}.  It is also straightforward 
to extend these models to higher dimensions, which will be discussed.

\subsection{Minimal model}
\label{subsec:321-5}

Let us consider a 5D supersymmetric gauge theory.  We consider that 
the fifth dimension, $y$, is compactified on an $S^1/Z_2$ orbifold, 
$0 \leq y \leq \pi R$, and that the gauge group in the bulk is $SU(5)$. 
We assume that the compactification radius is stabilized with $R^{-1} 
= O(10\!\sim\!100~{\rm TeV})$, which we typically take to be a factor 
of a few smaller than the fundamental scale $M_*$. 

The 5D gauge supermultiplet can be decomposed into a 4D $N=1$ vector 
superfield $V(A_\mu, \lambda)$ and a 4D $N=1$ chiral superfield 
$\Sigma(\sigma+iA_5, \lambda')$, where both $V$ and $\Sigma$ are in 
the adjoint representation of $SU(5)$.  We assume that these fields 
obey the following boundary conditions:
\begin{eqnarray}
  && V:\: \left( \begin{array}{ccc|cc}
    (+,+) & (+,+) & (+,+) & (-,+) & (-,+) \\ 
    (+,+) & (+,+) & (+,+) & (-,+) & (-,+) \\ 
    (+,+) & (+,+) & (+,+) & (-,+) & (-,+) \\ \hline
    (-,+) & (-,+) & (-,+) & (+,+) & (+,+) \\ 
    (-,+) & (-,+) & (-,+) & (+,+) & (+,+) 
  \end{array} \right),
\label{eq:bc-gauge-1} \\
  && \Sigma:\: \left( \begin{array}{ccc|cc}
    (-,-) & (-,-) & (-,-) & (+,-) & (+,-) \\ 
    (-,-) & (-,-) & (-,-) & (+,-) & (+,-) \\ 
    (-,-) & (-,-) & (-,-) & (+,-) & (+,-) \\ \hline
    (+,-) & (+,-) & (+,-) & (-,-) & (-,-) \\ 
    (+,-) & (+,-) & (+,-) & (-,-) & (-,-) 
  \end{array} \right),
\label{eq:bc-gauge-2}
\end{eqnarray}
where $+$ and $-$ represent Neumann and Dirichlet boundary conditions, 
respectively, and the first and second signs in parentheses represent 
boundary conditions at $y=0$ and $y=\pi R$, respectively.  This reduces 
the gauge symmetry at low energies to $SU(3) \times SU(2) \times U(1)$, 
which we identify as the standard model gauge group $SU(3)_C \times 
SU(2)_L \times U(1)_Y$ (321).  The active gauge group on the $y=0$ brane 
is 321, while that in all other places in the extra dimension is $SU(5)$ 
(see e.g.~\cite{Hall:2002ea}).  The typical mass scale for the KK towers 
is $R^{-1} = O(10\!\sim\!100~{\rm TeV})$, which we identify as $M_c$ 
in the previous sections.

The gauge couplings for the low-energy 321 gauge fields, $g_I$ ($I=1,2,3$), 
receive contributions both from the bulk and brane gauge couplings.  Here 
we assume that the brane contributions are small, giving only $O(1/8\pi^2)$ 
corrections to the inverse square couplings $1/g_I^2$.  This assumption 
is technically natural, and may be justified from physics above $M_*$. 
The 321 gauge couplings at the scale $M_c = 1/R$ are then given by the 
bulk contribution.  An important point is that this contribution is 
$SU(5)$ symmetric: $1/g_I^2 = \pi R/g_*^2$, where $g_*$ is the 5D $SU(5)$ 
gauge coupling.  Denoting the radion chiral superfield associated with 
the fifth dimension as $T$ and appropriately choosing the normalization 
for this field, we find that we exactly reproduce the first term of 
Eq.~(\ref{eq:gauge-kin}).  In particular, this explains the particular 
normalization for the coupling of $U(1)_Y$ gauge field to $T$.

We now gauge a $U(1)_R$ symmetry in this theory.  Since $U(1)_R$ does 
not commute with the 5D Lorentz symmetry ($U(1)_R$ contains the subgroup 
of $SU(2)_R$ in the 5D supersymmetry algebra that commutes with the 4D 
Lorentz symmetry), gauging it is associated with breaking of the 5D 
Lorentz symmetry to the 4D one.  Specifically, it will lead to a (small) 
nontrivial warping along the fifth dimension.  Here we assume that 
the resulting warping is small such that we can treat our 5D spacetime 
approximately flat, which can be the case depending on the explicit 5D 
setup.  (For an analysis of gauged $U(1)_R$ symmetries in 5D spacetime, 
see e.g.~\cite{Abe:2004nx}.)  The case with nontrivial warping will 
be discussed in section~\ref{subsec:warp}.

In general, a $U(1)_R$ symmetry gauged in our system could have 
anomalous matter content.  The resulting anomalies depend on the 
$U(1)_R$ charge assignment for matter, i.e. on what linear combination 
of the geometric $U(1)_R$ (the subgroup of $SU(2)_R$) and other 
``flavor'' $U(1)$'s we gauge as our $U(1)_R$ gauge symmetry.  To 
reproduce the effective theory of section~\ref{sec:framework} below 
$M_c$, we introduce three generations of quark and lepton supermultiplets, 
as well as a pair of Higgs-doublet and a singlet supermultiplets, 
in the bulk.  Each of these 5D supermultiplets (hypermultiplets) is 
decomposed into two 4D $N=1$ chiral superfields $\Phi(\phi,\psi)$ 
and $\Phi^c(\phi^c,\psi^c)$ with opposite gauge transformation 
properties.  The $U(1)_R$ charge assignment is then given by $\Phi(2/3)$ 
and $\Phi^c(4/3)$ for all matter and Higgs hypermultiplets (the 4D 
matter and Higgs superfields arise from $\Phi$; see below), leading 
to mixed $U(1)_R$ anomalies given in Eq.~(\ref{eq:U1R-anom}).  The 
$U(1)_R$ charges of $V$ and $\Sigma$ are zero.  The cancellation of 
these anomalies can then occur through the (generalized) Green-Schwarz 
mechanism with a modulus $M$, as discussed in section~\ref{sec:framework}. 
Assuming that $M$ is a brane-localized chiral superfield for simplicity, 
the $M$ field as well as the couplings of $M$ to the standard model 
gauge multiplets must be located on the $y=0$ brane (the 321 brane), 
since they do not respect $SU(5)$.%
\footnote{The case with a bulk $M$ field can also be considered 
with $M$ completed into an appropriate 5D (vector or tensor) 
supermultiplet; see e.g.~\cite{Dudas:2004ni}.}
The coefficients of these couplings are determined by the anomaly 
cancellation conditions.    This essentially reproduces the second 
term of Eq.~(\ref{eq:gauge-kin}) at the scale $M_c$, although it may 
have a slight modification arising from the existence of extra moduli 
and/or matter fields.  Such a modification may, in fact, be necessary 
to ensure the consistency of the effective field theory.  A more 
detailed analysis of the anomaly cancellation in higher dimensions, 
as well as the consistency of the higher dimensional theory, will 
be given later.

We now discuss matter and Higgs fields in more detail.  We impose 
boundary conditions on these fields such that each $SU(3)_C \times 
SU(2)_L \times U(1)_Y$ multiplet arises as the zero mode of a single 
5D hypermultiplet that transforms as a definite representation under 
$SU(5)$.  Consider, for example, a hypermultiplet $\{ {\cal D}, 
{\cal D}^c \}$ transforming as ${\bf 5}^*$ of $SU(5)$.  (In our 
notation, a conjugated field has the opposite transformation property 
from a non-conjugated field, and we specify the transformation property 
of a hypermultiplet by that of the non-conjugated chiral superfield; 
for instance, ${\cal D}$ and ${\cal D}^c$ transform as ${\bf 5}^*$ 
and ${\bf 5}$ under $SU(5)$, respectively.)  We choose the boundary 
conditions for this hypermultiplet as follows:
\begin{eqnarray}
  {\cal D}
    &=& {\cal D}_D^{(+,+)}({\bf 3}^*,{\bf 1})_{1/3}
      \oplus {\cal D}_L^{(-,+)}({\bf 1},{\bf 2})_{-1/2},
\label{eq:minimal-bc-D1} \\
  {\cal D}^c
    &=& {\cal D}_D^{c\,(-,-)}({\bf 3},{\bf 1})_{-1/3}
      \oplus {\cal D}_L^{c\,(+,-)}({\bf 1},{\bf 2})_{1/2}.
\label{eq:minimal-bc-D2}
\end{eqnarray}
The right-hand-side of these equations shows the decomposition 
of ${\cal D}$ and ${\cal D}^c$ into representations of 321 (in an 
obvious notation), as well as the boundary conditions imposed on each 
component (in the same notation as that in Eqs.~(\ref{eq:bc-gauge-1},%
~\ref{eq:bc-gauge-2})).  With these boundary conditions, the only 
massless state arising from $\{ {\cal D}, {\cal D}^c \}$ is the 
zero mode of ${\cal D}_D({\bf 3}^*,{\bf 1})_{1/3}$, which we identify 
as the low-energy down-type quark superfield $D$.  The other quark 
and lepton superfields are also obtained similarly.  Specifically, 
we introduce three generations of hypermultiplets $\{ {\cal Q}_i, 
{\cal Q}_i^c \}({\bf 10})$, $\{ {\cal U}_i, {\cal U}_i^c \}({\bf 10})$, 
$\{ {\cal D}_i, {\cal D}_i^c \}({\bf 5}^*)$, $\{ {\cal L}_i, {\cal L}_i^c 
\}({\bf 5}^*)$ and $\{ {\cal E}_i, {\cal E}_i^c \}({\bf 10})$ ($i=1,2,3$) 
for the quarks and leptons, obeying the following boundary conditions:
\begin{eqnarray}
  {\cal Q}
    &=& {\cal Q}_Q^{(+,+)}({\bf 3},{\bf 2})_{1/6}
      \oplus {\cal Q}_U^{(-,+)}({\bf 3}^*,{\bf 1})_{-2/3}
      \oplus {\cal Q}_E^{(-,+)}({\bf 1},{\bf 1})_{1},
\label{eq:minimal-bc-Q} \\
  {\cal U}
    &=& {\cal U}_Q^{(-,+)}({\bf 3},{\bf 2})_{1/6}
      \oplus {\cal U}_U^{(+,+)}({\bf 3}^*,{\bf 1})_{-2/3}
      \oplus {\cal U}_E^{(-,+)}({\bf 1},{\bf 1})_{1},
\label{eq:minimal-bc-U} \\
  {\cal D}
    &=& {\cal D}_D^{(+,+)}({\bf 3}^*,{\bf 1})_{1/3}
      \oplus {\cal D}_L^{(-,+)}({\bf 1},{\bf 2})_{-1/2},
\label{eq:minimal-bc-D} \\
  {\cal L}
    &=& {\cal L}_D^{(-,+)}({\bf 3}^*,{\bf 1})_{1/3}
      \oplus {\cal L}_L^{(+,+)}({\bf 1},{\bf 2})_{-1/2},
\label{eq:minimal-bc-L} \\
  {\cal E}
    &=& {\cal E}_Q^{(-,+)}({\bf 3},{\bf 2})_{1/6}
      \oplus {\cal E}_U^{(-,+)}({\bf 3}^*,{\bf 1})_{-2/3}
      \oplus {\cal E}_E^{(+,+)}({\bf 1},{\bf 1})_{1},
\label{eq:minimal-bc-E}
\end{eqnarray}
where we have omitted the generation index $i$.  (The boundary conditions 
for the conjugated fields are given by $+ \leftrightarrow -$, as in 
Eqs.~(\ref{eq:minimal-bc-D1},~\ref{eq:minimal-bc-D2}).)  The only 
massless states arising from these hypermultiplets are the zero modes of 
${\cal Q}_Q({\bf 3},{\bf 2})_{1/6}$, ${\cal U}_U({\bf 3}^*,{\bf 1})_{-2/3}$, 
${\cal D}_D({\bf 3}^*,{\bf 1})_{1/3}$, ${\cal L}_L({\bf 1},{\bf 2})_{-1/2}$ 
and ${\cal E}_E({\bf 1},{\bf 1})_{1}$, which we identify as $Q$, $U$, $D$, 
$L$ and $E$ in section~\ref{sec:framework}.  For the Higgs fields, 
we introduce $\{ {\cal H}, {\cal H}^c \}({\bf 5})$, $\{ \bar{\cal H}, 
\bar{\cal H}^c \}({\bf 5}^*)$ and $\{ {\cal S}, {\cal S}^c \}({\bf 1})$, 
obeying the boundary conditions:
\begin{eqnarray}
  {\cal H}
    &=& {\cal H}_C^{(-,+)}({\bf 3},{\bf 1})_{-1/3}
      \oplus {\cal H}_F^{(+,+)}({\bf 1},{\bf 2})_{1/2},
\label{eq:minimal-bc-H} \\
  \bar{\cal H}
    &=& \bar{\cal H}_C^{(-,+)}({\bf 3}^*,{\bf 1})_{1/3}
      \oplus \bar{\cal H}_F^{(+,+)}({\bf 1},{\bf 2})_{-1/2},
\label{eq:minimal-bc-Hbar} \\
  {\cal S}
    &=& {\cal S}_S^{(+,+)}({\bf 1},{\bf 1})_{0}.
\label{eq:minimal-bc-S}
\end{eqnarray}
(Again the boundary conditions for the conjugated fields are given by 
$+ \leftrightarrow -$.)  The massless states arise from the zero modes of 
${\cal H}_F({\bf 1},{\bf 2})_{1/2}$, $ \bar{\cal H}_F({\bf 1},{\bf 2})_{-1/2}$ 
and ${\cal S}_S({\bf 1},{\bf 1})_{0}$, which we identify as $H_u$, $H_d$ 
and $S$ in section~\ref{sec:framework}.  Note that the boundary conditions 
of Eqs.~(\ref{eq:bc-gauge-1},~\ref{eq:bc-gauge-2},~\ref{eq:minimal-bc-Q}~%
--~\ref{eq:minimal-bc-S}) can be imposed consistently with the interactions 
of the theory.%
\footnote{Note that the signs $\pm$ for the boundary conditions in 
these equations represent the Neumann/Dirichlet boundary conditions 
in the interval $y: [0, \pi R]$.  In the orbifold picture, the 
boundary conditions of e.g. Eq.~(\ref{eq:minimal-bc-U}) can be obtained 
effectively as follows.  We prepare a hypermultiplet obeying the 
boundary conditions ${\cal U} = {\cal U}_Q^{(-,+)}({\bf 3},{\bf 2})_{1/6} 
\oplus {\cal U}_U^{(+,+)}({\bf 3}^*,{\bf 1})_{-2/3} \oplus 
{\cal U}_E^{(+,+)}({\bf 1},{\bf 1})_{1}$, where the first and 
second signs in the parentheses represent transformation properties 
under the reflection $y \leftrightarrow -y$ and $(y-\pi R) \leftrightarrow 
-(y - \pi R)$, respectively.  We then introduce a 321-brane localized 
chiral superfield transforming as $({\bf 1},{\bf 1})_{-1}$ under 
321, and couple it to the ${\cal U}_E^{(+,+)}({\bf 1},{\bf 1})_{1}$ 
state from ${\cal U}$.  This reproduces the boundary conditions of 
Eq.~(\ref{eq:minimal-bc-U}) in the limit that this coupling (brane 
mass term) becomes large.  A similar (or more straightforward) 
construction also applies to the other multiplets.  The fact that the 
boundary conditions of Eqs.~(\ref{eq:bc-gauge-1},~\ref{eq:bc-gauge-2},~%
\ref{eq:minimal-bc-Q}~--~\ref{eq:minimal-bc-S}) can be reproduced in 
the orbifold picture by taking a consistent limit guarantees their 
consistency.}

A bulk hypermultiplet $\{ \Phi, \Phi^c \}$ can generically have a mass 
term in the bulk, which is written as 
\begin{equation}
  S = \int\!d^4x \int_0^{\pi R}\!\!dy 
      \int\! d^2\theta\, M_\Phi \Phi \Phi^c + {\rm h.c.},
\label{eq:bulk-mass}
\end{equation}
in the basis where the kinetic term is given by $S_{\rm kin} = \int\!d^4x 
\int\!dy\, [\int\!d^4\theta\, (\Phi^\dagger \Phi + \Phi^c \Phi^{c\dagger}) 
+ \{ \int\!d^2\theta\, \Phi^c \partial_y \Phi + {\rm h.c.} \}]$. 
The parameter $M_\Phi$ controls the wavefunction profile of the zero 
mode.  For $M_\Phi > 0$ ($< 0$) the wavefunction of a zero mode arising 
from $\Phi$ is localized to the $y=0$ ($y=\pi R$) brane; for $M_\Phi = 0$ 
it is flat.  (If a zero mode arises from $\Phi^c$, its wavefunction is 
localized to the $y=\pi R$ ($y=0$) brane for $M_\Phi > 0$ ($< 0$) and is 
flat for $M_\Phi = 0$.)  Explicit constraints on $M_\Phi$ in our theory 
depend on the detailed setup, e.g., on the source of supersymmetry 
breaking.  There is, however, one constraint that generically applies 
regardless of these details.  Suppose, for example, that we want to 
localize the zero mode of ${\cal D}_D({\bf 3}^*,{\bf 1})_{1/3}$ to 
the $y=\pi R$ brane by taking $M_{\cal D} \rightarrow -\infty$.  In 
this case, however, we find that the lightest KK state from ${\cal 
D}_L({\bf 1},{\bf 2})_{-1/2}$ and ${\cal D}_L^c({\bf 1},{\bf 2})_{1/2}$ 
becomes exponentially light, with the former (latter) degrees of freedom 
localized to the $y=\pi R$ ($y=0$) brane.  This is, in fact, expected 
because the ${\cal D}_D$ state is localized to the $y=\pi R$ brane, 
where the active gauge group is $SU(5)$, so that it locally requires 
an $SU(5)$ partner, which is provided by the ${\cal D}_L$ state. 
Since the ${\cal D}_L$ state is massive in 4D (due to the boundary 
conditions), it must be in a vector-like representation, hence the 
existence of ${\cal D}_L^c$ localized at $y=0$.  Since any extra 
vector-like state is not observed in nature, this gives a constraint 
on $M_{\cal D}$ from below.  Applying similar considerations also 
to the other multiplets, we find
\begin{equation}
  M_\Phi R \simgt -(1\!\sim\!2),
\label{eq:const-M}
\end{equation}
for $\Phi = {\cal Q}_i, {\cal U}_i, {\cal D}_i, {\cal L}_i, {\cal E}_i, 
{\cal H}$ and $\bar{\cal H}$, implying that the wavefunctions of the 
low-energy states $Q_i$, $U_i$, $D_i$, $L_i$, $E_i$, $H_u$ and $H_d$ should 
not be strongly localized to the $y=\pi R$ brane (the $SU(5)$ brane).%
\footnote{One may think that for a $\{ {\cal D}, {\cal D}^c \}$ multiplet 
with $M_{\cal D} \rightarrow -\infty$, we can introduce a 321-brane 
localized field $L'({\bf 1},{\bf 2})_{-1/2}$ and couple it to ${\cal D}_L^c$ 
on the 321 brane, leading to the low-energy states ${\cal D}_D$ and 
${\cal D}_L$ (with a slight mixture from $L'$), which may be identified 
as $D$ and $L$.  (The $\{ {\cal L}, {\cal L}^c \}$ multiplet should 
then be eliminated.)  In fact, this construction can work for a ${\bf 5}^*$ 
($\supset D+L$) state, although a similar construction for a ${\bf 10}$ 
($\supset Q+U+E$) state does not because it would lead to rapid proton 
decay caused by an exchange of the bulk $SU(5)$ gauge boson.  This opens 
a possibility in which (some of) the ${\bf 5}^*$ states are strongly 
(or exactly) localized to the $SU(5)$ brane.}

With the structure for the matter and Higgs sectors described above, no 
rapid proton decay is induced by an exchange of the bulk $SU(5)$ gauge 
boson, whose mass is only of order $1/R \approx (10\!\sim\!100)~{\rm TeV}$. 
This is because quarks and leptons that would be unified into a single 
$SU(5)$ representation in standard grand unified theories now arise from 
different $SU(5)$ multiplets in the bulk.  (Note that this preserves an 
$SU(5)$ understanding of the quark and lepton quantum numbers, especially 
quantization of $U(1)_Y$ hypercharges.)  The colored-Higgs KK states with 
masses of $O(10\!\sim\!100~{\rm TeV})$ do not induce proton decay either, 
because of the special form of the mass matrices for these states, dictated 
by higher dimensional gauge invariance.  (See Ref.~\cite{Hall:2002ea} for 
details.)  Possible tree-level proton decay operators may be forbidden 
by imposing an appropriate discrete (gauge) symmetry or if the quarks 
and leptons are separated in an extra dimension orthogonal to the 
dimension $y$ (see section~\ref{subsec:higher-d}).

The superpotential interactions arise from the $y=0$ (321) and/or 
$y=\pi R$ ($SU(5)$) brane(s).  Due to the gauged $U(1)_R$ symmetry, 
these interactions must be cubic in fields.  Locating them on the 
321 brane for simplicity, we obtain
\begin{equation}
  S = \int\!d^4x \int_0^{\pi R}\!\!dy\,\, \delta(y)\!
      \int\! d^2\theta\, \Bigl( {\cal Q} {\cal U} {\cal H} 
      + {\cal Q} {\cal D} \bar{\cal H} + {\cal L} {\cal E} \bar{\cal H} 
      + {\cal S} {\cal H} \bar{\cal H} + {\cal S}^3 \Bigr) + {\rm h.c.},
\label{eq:Yukawa-321-5}
\end{equation}
where we have omitted generation indices as well as (dimensionful) 
coefficients, and we have assumed the standard matter parity 
($R$ parity).  These interactions reproduce the interactions of 
Eq.~(\ref{eq:W}) below $M_c$.  (Small neutrino masses can be generated 
by introducing a brane or bulk right-handed neutrinos ${\cal N}$ 
together with a superpotential term of the form ${\cal L} {\cal N} 
{\cal H}$; see section~\ref{subsec:others}.)  Note that there is 
no unwanted unified mass relation between the quarks and leptons, 
since different 321 multiplets come from different bulk multiplets. 

We now discuss anomaly cancellation for the $U(1)_R$ symmetry 
in the present theory in more detail.  In general, when one performs 
a $U(1)_R$ transformation, variations of the Lagrangian caused by the 
anomalous matter content are confined to the branes at $y=0$ and 
$y=\pi R$~\cite{Arkani-Hamed:2001is}.  In our theory, the generated 
anomalies take the form
\begin{equation}
  \left( \begin{array}{c}
    A_1^{(0)} \\ A_2^{(0)} \\ A_3^{(0)} 
  \end{array} \right)
  = \left( \begin{array}{c}
    -\frac{11}{5}-a \\ -\frac{1}{3}-a \\ 1-a 
  \end{array} \right),
\qquad
  \left( \begin{array}{c}
    A_1^{(\pi)} \\ A_2^{(\pi)} \\ A_3^{(\pi)} 
  \end{array} \right)
  = \left( \begin{array}{c}
    a \\ a \\ a 
  \end{array} \right),
\label{eq:U1R-anom-5D}
\end{equation}
where $A_I^{(0)}$ and $A_I^{(\pi)}$ ($I=1,2,3$) represent the mixed 
$U(1)_R$-321 anomalies located at the $y=0$ and $y=\pi R$ branes, 
respectively.  The constant $a$ is given by $a = -5/12$ in the model 
described above, although the value of $a$ changes in general if we 
locate (some of) the quark, lepton and Higgs multiplets strictly on 
the 321 brane.  The mixed anomalies of Eq.~(\ref{eq:U1R-anom-5D}) 
can be canceled by a combination of the Green-Schwarz mechanism and 
anomaly transfer in the bulk.  By introducing a bulk Chern-Simons term 
with an appropriate coefficient, we can ``transfer'' the mixed anomalies 
from $y=\pi R$ to $y=0$ by an amount of $a$.  The rest of the anomalies 
can then be canceled by introducing the terms
\begin{equation}
  S = \int\!d^4x \int_0^{\pi R}\!\!dy\,\, \delta(y)
    \Biggl\{ -\sum_{I=1,2,3} \frac{\zeta_I}{c} \int\!d^2\theta\, 
    M\, {\cal W}_I^{a\alpha} {\cal W}_{I\alpha}^a + {\rm h.c.} \Biggr\},
\label{eq:gen-GS-5D}
\end{equation}
with the coefficients chosen to be $\zeta_I = A_I \equiv A_I^{(0)} 
+ A_I^{(\pi)}$.  Here, $\int_0^\epsilon \delta(y)\, dy \equiv 1$ for 
$\epsilon > 0$, and $M$ transforms as $M \rightarrow M + i \alpha 
c /16\pi^2$ under $U(1)_R$.  After integrating out the fifth dimension, 
the interactions of Eq.~(\ref{eq:gen-GS-5D}) lead exactly to the second 
term of Eq.~(\ref{eq:gauge-kin}) at the scale $M_c$.  (The first term 
arises from the bulk gauge kinetic term.)

A (potential) problem with the setup just described is that the 
interactions of Eq.~(\ref{eq:gen-GS-5D}) give a negative brane-localized 
kinetic term for $SU(3)_C$ after the modulus $M$ obtains the required 
VEV of Eq.~(\ref{eq:corresp-2}).  (Note that $\langle M \rangle/c > 0$ 
and $\zeta_3 = 1$.)  While the zero mode of the $SU(3)_C$ gauge field 
has a positive gauge kinetic term, the negative brane kinetic term 
could cause problems in processes involving higher KK states.  Suppose 
that the coefficients of the bulk and brane kinetic terms for a bulk 
gauge field are given by $1/g_5^2$ and $1/\tilde{g}^2$, respectively. 
(For the $SU(3)_C$ gauge field considered here, $1/g_5^2 = 1/g_*^2$ 
and $1/\tilde{g}^2 = -4\langle M \rangle/c$.)  We then find that for 
$1/\tilde{g}^2 < 0$, the KK decomposition leads to a mode that has 
a negative kinetic term (ghost), whose ``mass'' $\mu_0$ ($>0$) is 
given by the solution to
\begin{equation}
  \tanh(\pi R \mu_0) = -\frac{g_5^2 \mu_0}{\tilde{g}^2}.
\label{eq:mu0}
\end{equation}
In order for the 5D effective field theory to be consistent, this 
mass must be larger than the cutoff scale: $\mu_0 \simgt M_*$, 
leading to the condition $1/\tilde{g}^2 \simgt -1/g_5^2 M_*$.  For 
the case of $SU(3)_C$, however, the values of $1/\tilde{g}^2$ and 
$1/g_5^2$ are determined by the phenomenological requirements of 
Eqs.~(\ref{eq:corresp-1},~\ref{eq:corresp-2}) as $1/\tilde{g}^2 
\simeq -1$ and $\pi R/g_5^2 \simeq 2$.  This yields $M_* \simlt 
2/\pi R \approx 1/R$, implying that the cutoff scale of the 5D 
theory should be at or below the scale of the masses of the first 
KK excitations.  This clearly casts doubt on the viability of the 
5D theory as a theory describing physics ``above $M_c$.''

There are essentially two approaches we could take to deal with this 
issue.  One is to consider that the size of the extra dimension, $\pi R$, 
is in fact not much larger than the cutoff length $M_*^{-1}$.  In this 
case, the 5D theory described above may not be a fully viable effective 
field theory.  However, we can still take the view that it suggests 
the basic structure, e.g. the gauge symmetry structure and matter 
content, of the fundamental theory at $M_*$, e.g. string theory.  This 
is an interesting proposal for future string model-building.  The other 
approach is to consider that the problem arose because of the particular 
(too minimal) structure of the model described above, and that we can 
(slightly) modify the theory so that it does not suffer from the problem. 
Below we take this latter approach and find ways to avoid the 
problem within effective field theory.

A simple way of avoiding the sizable negative gauge kinetic term for 
$SU(3)_C$ on the $y=0$ brane is to introduce another modulus $M'$, 
which is localized on the $y=\pi R$ brane and has the interaction
\begin{equation}
  S = \int\!d^4x \int_0^{\pi R}\!\!dy\,\, \delta(y-\pi R)
    \Biggl\{ - \frac{\zeta'}{c'} \int\!d^2\theta\, 
    M'\, {\cal W}^{a\alpha} {\cal W}_{\alpha}^a + {\rm h.c.} \Biggr\},
\label{eq:M'-5D}
\end{equation}
where $\zeta'$ and $c'$ are real constants, ${\cal W}^{a\alpha}$ is 
the field strength superfield for $SU(5)$, which contains 321 as a 
subgroup, and the field $M'$ transforms as $M' \rightarrow M' + i \alpha 
c' /16\pi^2$ under $U(1)_R$.  Note that the active gauge group on the 
$y=\pi R$ brane is $SU(5)$, so that the coefficient $\zeta'$ is universal 
for 321.  In this case, the coefficient for the bulk Chern-Simons term 
should be chosen such that mixed anomalies of the amount $a-\zeta'$ 
are transfered from $y=\pi R$ to $y=0$, and the coefficients $\zeta_I$ 
in Eq.~(\ref{eq:gen-GS-5D}) chosen as
\begin{equation}
  \zeta_I = A_I - \zeta',
\label{eq:zeta_I}
\end{equation}
where $(A_1,A_2,A_3)=(-11/5,-1/3,1)$.  Now, let us consider, for example, 
that $\zeta' = 1$.  In this case the coefficients of the 321-brane 
localized interactions take the values $(\zeta_1,\zeta_2,\zeta_3) 
= (-16/5,-4/3,0)$, so that the VEV of $\langle M \rangle/c > 0$ does 
not lead to a negative brane kinetic term for $SU(3)_C$, $SU(2)_L$ 
or $U(1)_Y$.  In fact, assuming that $\langle M' \rangle/c' \simlt 
1/16\pi^2$ (not necessarily $|\langle M' \rangle/c'| \simlt 1/16\pi^2$), 
we can make our 5D theory a viable effective field theory in a 
(moderately) large energy interval above $M_c \approx 1/R$.  To 
reproduce the observed gauge couplings, we must have
\begin{eqnarray}
  && \frac{\langle M \rangle}{c} \simeq 
  \frac{3}{32\pi^2}\ln\frac{M_U}{M_*} \simeq 0.25,
\label{eq:req-M}
\\
  && \frac{\pi R}{g_*^2} - \frac{4\zeta'}{c'}\langle M' \rangle
  = \langle T \rangle - 4\frac{\langle M' \rangle}{c'}
  \simeq 1.
\label{eq:req-T}
\end{eqnarray}
By choosing the 5D $SU(5)$ gauge coupling to be strong at the scale 
$M_*$, i.e. $1/g_*^2 \approx C M_*/16\pi^3$ with $C \simeq 5$ a group 
theoretical factor, we find that $M_* R$ can be as large as $\approx 30$ 
for $|\langle M' \rangle/c'| \ll 1$.  These choices of parameters do 
not disturb any of the arguments before, e.g. technical naturalness 
for the smallness of the tree-level brane gauge kinetic operators.  The 
fields $M$ and $M'$ can be stabilized easily with the desired values 
of $\langle M \rangle/c$ and $\langle M' \rangle/c'$ along the lines 
of section~\ref{subsec:M-stab}.  For instance, we can consider two 
supersymmetric $SU(2)$ gauge sectors each localized on the $y=0$ and 
$y=\pi R$ branes, which are responsible for the stabilizations of $M$ 
and $M'$, respectively. 

An alternative way of avoiding the problem is to introduce extra 
matter fields that are vector-like under 321 and obtain masses through 
$U(1)_R$ breaking.  Let us, for example, introduce chiral superfields 
$\Phi({\bf 5}) + \bar{\Phi}({\bf 5}^*)$ on the $y=\pi R$ brane which 
have a vanishing $U(1)_R$ charge.  Here, the numbers in parentheses 
represent the transformation properties under $SU(5)$.  In this case, 
these fields produce the mixed anomalies of $-1$ localized on the 
$y=\pi R$ brane, so that $A_I^{(\pi)}$ in Eq.~(\ref{eq:U1R-anom-5D}) 
are replaced as $A_I^{(\pi)} = a \rightarrow a-1$.  The anomaly transfer 
by a Chern-Simons term should then be $a-1$, and the coefficients $\zeta_I$ 
in Eq.~(\ref{eq:gen-GS-5D}) become $\zeta_I = A_I - 1 \leq 0$, avoiding 
the problem.  The required values of $\langle T \rangle$ and $\langle 
M \rangle$ are given by Eqs.~(\ref{eq:req-M},~\ref{eq:req-T}) (with 
$\langle M' \rangle$ set to zero), and we find, following the argument 
below Eqs.~(\ref{eq:req-M},~\ref{eq:req-T}), that there can be a 
(moderately) large energy interval up to a factor of $M_* R \approx 30$ 
in which the effective 5D field theory is applicable.  A mass for the 
$\Phi$ and $\bar{\Phi}$ states of order the weak scale or somewhat larger 
can be generated through the K\"ahler potential term $\int\!d^4\theta\, 
C^\dagger C \Phi \bar{\Phi}$ on the $y=\pi R$ brane, where $C$ represents 
the chiral compensator field.  (This requires that supersymmetry is broken 
in the bulk of the gravitational dimensions, in which case $F_C$ is of 
order the weak scale or somewhat larger; see section~\ref{subsec:susy-br}.) 
To preserve the successful prediction for the gauge couplings, the 
absence of similar vector-like states on the 321 brane which do not 
fill a complete $SU(5)$ multiplet must be assumed.  A similar comment 
also applies to states that obtain masses though the VEV of $\phi$, 
the field absorbing the large Fayet-Iliopoulos term of $U(1)_R$.

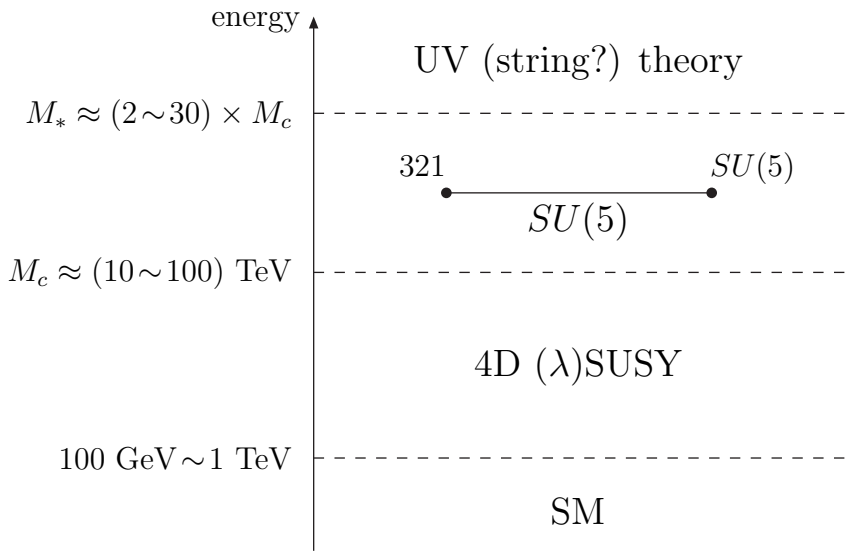
\begin{figure}[t]
\begin{center} 
\begin{picture}(275,205)(-50,0)
  \LongArrow(0,0)(0,200) \Text(-5,200)[r]{energy}
  \Text(100,185)[]{\large UV (string?) theory}
  \Text(-8,165)[r]{$M_* \approx (2\!\sim\!30) \times M_c$}
  \DashLine(0,165)(200,165){4}
  \Line(50,135)(150,135) \Text(100,132)[t]{\large $SU(5)$}
  \Vertex(50,135){2}  \Text(50,142)[br]{321}
  \Vertex(150,135){2} \Text(150,142)[bl]{$SU(5)$}
  \Text(-8,105)[r]{$M_c \approx (10\!\sim\!100)~{\rm TeV}$}
  \DashLine(0,105)(200,105){4}
  \Text(100,70)[]{\large 4D ($\lambda$)SUSY}
  \Text(-8,35)[r]{$100~{\rm GeV}\!\sim\!1~{\rm TeV}$}
  \DashLine(0,35)(200,35){4}
  \Text(100,15)[]{\large SM}
\end{picture}
\caption{The schematic picture of our minimal theory.  The standard 
 model (SM) is the effective theory up to a scale of $100~{\rm 
 GeV}\!\sim\!1~{\rm TeV}$, where it is replaced by a 4D ($\lambda$)SUSY 
 model: a 4D $N=1$ supersymmetric standard model with the superpotential 
 given by Eq.~(\ref{eq:W}) (in the minimal case).  This model is further 
 replaced at $M_c \approx (10\!\sim\!100)~{\rm TeV}$ by the minimal 
 5D $SU(5)$ theory described in the text, which is the effective theory 
 for the next factor of $(2\!\sim\!30)$.  Finally, at the scale $M_* 
 \approx (2\!\sim\!30) \times M_c$, the theory is embedded into a 
 fundamental ultraviolet (UV) theory, which may be string theory.}
\label{fig:picture}
\end{center}
\end{figure}

We emphasize that the mechanisms presented above for avoiding a sizable 
negative brane kinetic term for $SU(3)_C$ are actually simple -- much 
simpler than how they might naively look.  We simply assume that the mixed 
$U(1)_R$ anomalies are canceled by a combination of the $M$ field and 
the $M'$ field (or extra vector-like states).  The interactions (quantum 
numbers) of $M'$ (vector-like states) are universal for 321 because 
of the location of the field(s), so that the successful supersymmetric 
prediction for the low-energy gauge couplings is preserved.  Note that 
this requires some reinterpretations of the formulae given in the previous 
sections, for example $F_T$ in Eqs.~(\ref{eq:M_I},~\ref{eq:gaugino-unif}) 
should be replaced by $F_T + (4\zeta'/c)F_M$, but the essential physics 
is unchanged.  The size of the 5D energy interval $M_* R$ takes a value 
between a factor of a few and $\approx 30$.  We thus arrive at the 
picture given in Fig.~\ref{fig:picture}. 
This completes our discussion on the basic construction of the model 
reproducing the effective theory of section~\ref{sec:framework} 
below $M_c$.

Let us now discuss supersymmetry breaking and its implications in 
this theory.  As discussed in section~\ref{subsec:susy-br}, one of 
the natural possibilities is that the $T$ field, the radion supermultiplet 
associated with the dimension $y$, obtains a nonvanishing VEV in the 
auxiliary component, $F_T = O(100~{\rm GeV})$.  This induces supersymmetry 
breaking masses for the gauginos as well as the bulk scalar fields. 
(For flat spacetime, this is equivalent to the Scherk-Schwarz 
mechanism~\cite{Scherk:1978ta}.  For earlier work on Scherk-Schwarz 
supersymmetry breaking, see e.g.~\cite{Antoniadis:1990ew,Pomarol:1998sd,%
Barbieri:2000vh}.)  Since the generated scalar masses and scalar trilinear 
interactions depend on bulk mass parameters $M_\Phi$, this generically 
introduces the supersymmetric flavor problem.  One way to avoid this 
problem is to assume that the bulk masses are flavor universal, which 
may be the result of some flavor symmetry.  This can lead to interesting 
phenomenology, with a variety of spectra for the squarks and sleptons 
depending on arbitrary bulk mass parameters.  Another possible way, 
which we focus on below, is to strongly localize the quark and lepton 
multiplets to a brane, since then the generated tree-level supersymmetry 
breaking masses for the squarks and sleptons are exponentially suppressed. 
Flavor universal squark and slepton masses can be generated by gauge 
loops through the gaugino masses (approximately flavor universal 
contributions can also come from the $U(1)_R$ $D$-term VEV), and scalar 
trilinear interactions proportional to the Yukawa matrices are also 
generated by gauge loops as well as by the tree-level contribution 
through the Higgs fields.  Since we have a constraint on $M_\Phi$ 
in Eq.~(\ref{eq:const-M}), we should then take
\begin{equation}
  M_{{\cal Q}_i}, M_{{\cal U}_i}, M_{{\cal D}_i}, 
    M_{{\cal L}_i}, M_{{\cal E}_i} 
  \gg \frac{1}{\pi R},
\label{eq:localization}
\end{equation}
implying that the low-energy $Q_i$, $U_i$, $D_i$, $L_i$ and $E_i$ 
states are localized to the 321 brane.%
\footnote{One of the main reasons we did not localize these states 
strictly on the 321 brane from the beginning is that we would then 
lose the $SU(5)$ understanding of the matter quantum numbers in 
the 5D effective theory.  The correct quantum numbers, however, 
can arise naturally if the fundamental theory is higher dimensional 
and has a larger gauge group, as in Ref.~\cite{Hall:2002qw}.}
The required amount of localization, however, is not very strong; 
$M R \simgt 2$ is enough for the first two generations, and the degree 
of localization can be even milder for the third generation.  For the 
Higgs and $S$ fields, there are no strong constraints on their bulk 
masses from flavor violating processes.  The constraints, however, 
may arise from electroweak symmetry breaking, depending on details 
of the setup, for example to avoid too large volume suppressions for 
the low-energy $\lambda$ and $\kappa$ couplings in Eq.~(\ref{eq:W}) 
and/or to have a sufficiently large scalar trilinear coupling 
between the $S$ and Higgs fields.

\begin{figure}[t]
\begin{center}
    \input{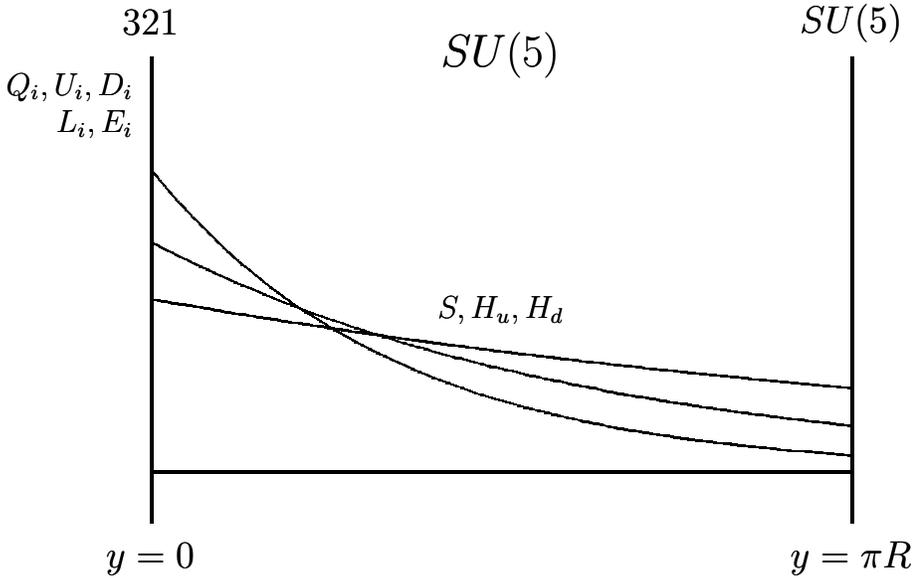}
\caption{The schematic picture of the minimal model described in the 
 text.  The quark and lepton supermultiplets are localized (strongly) 
 to the $y=0$ brane.  The wavefunction profiles for the $S$ and Higgs 
 fields are depicted arbitrarily.}
\label{fig:321-5}
\end{center}
\end{figure}
In Fig.~\ref{fig:321-5} we present a schematic picture of the model 
described here.  The wavefunction profiles for the $S$ and Higgs fields 
are depicted arbitrarily.  The structure of this theory is somewhat 
similar to that in Ref.~\cite{Barbieri:2001yz}, although we now have 
a low compactification scale of $1/R \approx (10\!\sim\!100)~{\rm TeV}$ 
and the gauged $R$ symmetry.  These two ingredients provide extra 
constraints on the location of the matter and Higgs fields, as well 
as on the form of the superpotential, given in Eqs.~(\ref{eq:const-M},~%
\ref{eq:Yukawa-321-5}).  Supersymmetry breaking masses also show 
a characteristic pattern.  Motivated by suppressions of flavor 
changing neutral currents, let us consider the situation in which 
the source of supersymmetry breaking resides in nonvanishing $F_T$, 
$D_R$ and anomaly mediation (see section~\ref{subsec:susy-br}). 
In this case, soft supersymmetry breaking masses at the scale 
$M_c = 1/R = O(10\!\sim\!100~{\rm TeV})$ are given as follows. 
For the gaugino masses $M_I$ ($I=1,2,3$), we have
\begin{equation}
  M_I = \frac{g_I^2}{g_U^{\prime 2}} \hat{m}_T 
    + \frac{b_I g_I^2}{16\pi^2} \hat{m}_C,
\label{eq:gaugino}
\end{equation}
where $g_I$ are the 321 gauge couplings at $M_c$, $g_U^{\prime 2} 
\equiv g_*^2/\pi R$ with $g_*$ the 5D $SU(5)$ gauge coupling, and 
$b_I$ are the beta-function coefficients defined by $d(1/g_I^2)/d\ln\mu_R 
= -b_I/8\pi^2$.  (The value of $g'_U$ would agree with the unified gauge 
coupling in the conventional supersymmetric desert, $g_U \simeq 0.7$, 
for $\zeta'=0$.)  The mass parameters $\hat{m}_T$ and $\hat{m}_C$ are 
given by
\begin{equation}
  \hat{m}_T \equiv -\frac{g_U^{\prime 2}}{2}F_T,
\qquad
  \hat{m}_C \equiv -F_C.
\label{eq:mT-mC}
\end{equation}
The two terms in Eq.~(\ref{eq:gaugino}) give comparable contributions 
for $\hat{m}_T \approx \hat{m}_C/8\pi^2$ ($\approx O(100~{\rm GeV})$); 
otherwise, one dominates the other.  The scalar trilinear interactions, 
defined generally by ${\cal L}_{\rm soft} = -\sum_{A,B,C} (a_{ABC}/6) 
\phi_A \phi_B \phi_C + {\rm h.c.}$, are given by
\begin{equation}
  a_{ABC} = -y_{ABC} \Bigl\{ (a_{\Phi_A}+a_{\Phi_B}+a_{\Phi_C})\, \hat{m}_T 
    + (\gamma_{\Phi_A}+\gamma_{\Phi_B}+\gamma_{\Phi_C})\, \hat{m}_C \Bigr\},
\label{eq:a_ABC}
\end{equation}
where $y_{ABC}$ are the Yukawa couplings $W = \sum_{A,B,C} (y_{ABC}/6) 
\Phi_A \Phi_B \Phi_C$, with $\Phi_A$ and $\phi_A$ representing a generic 
4D chiral superfield and its scalar component, respectively.%
\footnote{Our sign convention for the soft supersymmetry breaking parameters 
agrees with that of SUSY Les Houches Accord~\cite{Skands:2003cj}.}
The coefficient $a_\Phi$ is given in flat space by
\begin{equation}
  a_\Phi = \frac{2\pi R M_\varphi}{e^{2\pi R M_\varphi}-1},
\label{eq:a_phi}
\end{equation}
where $M_\varphi$ is the bulk mass of the hypermultiplet $\{ \varphi, 
\varphi^c \}$ giving $\Phi$ as the zero mode of $\varphi$, and $\gamma_\Phi$ 
is the anomalous dimension of $\Phi$, defined by $d\ln Z_\Phi/d\ln\mu_R = 
-2\gamma_\Phi$ with $Z_\Phi$ the wavefunction renormalization for $\Phi$.%
\footnote{Note that the 4D Yukawa couplings $y_{ABC}$ also receive 
suppressions with factors $z_{\Phi_A} z_{\Phi_B} z_{\Phi_C}$, where 
$z_\Phi = (2 M_\varphi/(1-e^{-2\pi R M_\varphi})M_*)^{1/2}$.  The 
suppression factors, however, could differ if these couplings receive 
nonnegligible contributions from the brane couplings located at $y=\pi R$. 
The expressions for the scalar trilinear couplings $a_{ABC}$ also change 
in this case from that in Eq.~(\ref{eq:a_ABC}).  This can happen, for 
example, for the couplings $\lambda$ and $\kappa$, practically making 
the corresponding scalar trilinear interactions, $a_\lambda$ and 
$a_\kappa$, independent free parameters.} 
The soft supersymmetry breaking scalar mass squared $m_\phi^2$ for 
a 4D chiral superfield $\Phi$ is given by
\begin{eqnarray}
  m_\phi^2 &=& - \gamma_\Phi \hat{m}_D^2 + c_\Phi |\hat{m}_T|^2
    + \frac{1}{2} \frac{d\gamma_\Phi}{d\ln\mu_R} |\hat{m}_C|^2
\nonumber\\
  && {} + \Biggl\{ \Biggl( \sum_{y_{ABC}} |y_{ABC}|^2 
      (a_{\Phi_A}+a_{\Phi_B}+a_{\Phi_C}) 
      \frac{\partial\gamma_\Phi}{\partial|y_{ABC}|^2}
      + \sum_{g_I} \frac{g_I^4}{g_U^{\prime 2}} 
        \frac{\partial\gamma_\Phi}{\partial g_I^2} \Biggr) 
      \hat{m}_T \hat{m}_C^* + {\rm h.c.} \Biggr\},
\label{eq:m2}
\end{eqnarray}
where the mass parameter $\hat{m}_D$ is given by
\begin{equation}
  \hat{m}_D^2 = -\frac{2}{3} D_R,
\label{eq:mD2}
\end{equation}
which is positive and can take a value of order the weak scale, or 
somewhat larger,%
\footnote{In the presence of a $U(1)_R$-$U(1)_Y$ kinetic mixing term on the 
321 brane, the scalar squared masses receive contributions proportional to 
their $U(1)_Y$ hypercharges.  Writing the gauge kinetic terms as $\int\!d^2\theta\, 
\{ (1/4g_R^2) {\cal W}_R^\alpha {\cal W}_{R\alpha} + (1/4g_1^2) {\cal W}_1^\alpha 
{\cal W}_{1\alpha} + (\epsilon/2) {\cal W}_R^\alpha {\cal W}_{1\alpha} \} 
+ {\rm h.c.}$ in 4D, the first term of Eq.~(\ref{eq:m2}) is then modified to 
$(-\gamma_\Phi - 3\epsilon g_1^2 Y_\Phi/2) \hat{m}_D^2$, where $Y_\Phi$ 
is the $U(1)_Y$ hypercharge of the chiral superfield $\Phi$ in the $SU(5)$ 
normalization.  Although it is (technically) natural to have a small value 
for $\epsilon$, it can also be of $O(1)$ without contradicting phenomenological 
constraints. \label{ft:kin-mix}}
and the coefficient $c_\Phi$ is given (in the flat space limit) by
\begin{equation}
  c_\Phi = \left(\frac{\pi R M_\varphi}{\sinh(\pi R M_\varphi)}\right)^2.
\label{eq:c_phi}
\end{equation}
The last term in the right-hand-side of Eq.~(\ref{eq:m2}) is the interference 
term between $F_T$ and $F_C$~\cite{Choi:2004sx}, and we have assumed the 
absence of direct couplings between the observable sector fields and $\phi$, 
the field absorbing the Fayet-Iliopoulos term of $U(1)_R$.  More explicit 
expressions for the scalar trilinear interactions, $a_{ABC}$, and 
scalar squared masses, $m_\phi^2$, in the present setup are given 
in Appendix~\ref{app:soft}.

The soft supersymmetry breaking masses derived above, 
Eqs.~(\ref{eq:gaugino},~\ref{eq:a_ABC},~\ref{eq:m2}), do not lead 
to the supersymmetric flavor problem, as long as the conditions of 
Eq.~(\ref{eq:localization}) are satisfied (at least for the first two 
generations).  They do not lead to the supersymmetric $CP$ problem either, 
if the complex phases of $\hat{m}_T$ and $\hat{m}_C$ ($F_T$ and $F_C$) 
are aligned, which is indeed the case if $T$ is stabilized through 
a single gaugino condensation~\cite{Choi:2004sx,Endo:2005uy}, or if 
one of $\hat{m}_T$ and $\hat{m}_C$ dominates the other.  (All the soft 
supersymmetry breaking masses, as well as the couplings $\lambda$ and 
$\kappa$, can be made real by choosing the appropriate phase convention 
for $S$ and $H_u H_d$ and by using the appropriate $R$-rotation.) 
These soft supersymmetry breaking masses show a variety of interesting 
patterns.  For example, if the contribution from anomaly mediation is 
small, $\hat{m}_C/8\pi^2 \ll \hat{m}_T, \hat{m}_D/4\pi$, we find that the 
gaugino masses satisfy the standard ``unified mass relation'', $M_I \propto 
g_I^2$, while the squark and slepton masses take the form $m_{\tilde{F}}^2 
\simeq -\gamma_F \hat{m}_D^2$, where $F = Q,U,D,L,E$ (at least for 
the first two generations).%
\footnote{This is the case if the Fayet-Iliopoulos term for $U(1)_R$ 
is not absorbed.}
The Higgs squared masses are arbitrary, and there are 
scalar trilinear terms ($A$ terms) proportional to the 
Yukawa matrices. Note that these masses are given at the scale 
$M_c = O(10\!\sim\!100~{\rm TeV})$, only a few orders of magnitude 
above the weak scale.  In the case that $F_T$ is suppressed, on the 
other hand, we find that the spectrum is given by the sum of the anomaly 
mediated contributions and the scalar masses from the $U(1)_R$ $D$-term 
VEV, again at the scale $M_c = O(10\!\sim\!100~{\rm TeV})$.  This can 
happen if there is a supersymmetry breaking field $Z$ in the gravitational 
bulk, which does not directly couple to the MSSM states.  In general, the 
superparticle spectra derived in Eqs.~(\ref{eq:gaugino},~\ref{eq:a_ABC},%
~\ref{eq:m2}) have a very rich structure, including the possibility 
of nonuniversality in the third generation sfermion masses caused by 
nontrivial profiles of the wavefunctions of these states in the extra 
dimension.  In addition, the gaugino masses may also have a contribution 
from $F_M$, which affects the scale of effective gaugino mass unification. 
We can even consider some interesting variations of the model.  For example, 
we can change the boundary conditions of ${\cal S}$ (and/or ${\cal H}_F$, 
$\bar{\cal H}_F$) to $(+,-)$ or $(-,+)$ (only $(+,-)$ is available for 
${\cal H}_F$, $\bar{\cal H}_F$).  In this case, the tree-level contribution 
to the supersymmetry breaking mass squared $m_{\cal S}^2$ (and/or 
$m_{\cal H}^2$, $m_{\bar{\cal H}}^2$) can be negative in a certain 
parameter region.  A detailed study of electroweak symmetry breaking 
for some of these spectra will be given in Ref.~\cite{NP}.

We finally discuss physics associated with the KK states, which have 
masses of order $M_c = 1/R = O(10\!\sim\!100~{\rm TeV})$.  These 
states form multiplets of 4D $N=2$ supersymmetry, with small mass 
splittings inside each supermultiplet due to $F_T$.  Interesting 
quantities among others are masses of the lightest KK excitations for 
the gauge fields.  These are determined independently of bulk mass 
parameters, and thus provide relatively model-independent predictions. 
We find that the masses of the KK gauge states associated with 
$SU(3)_C$, $SU(2)_L$, $U(1)_Y$ and $SU(5)/321$ are given by $M'_3 
= (g_3^2/g_U^{\prime 2})R^{-1}$, $M'_2 = (g_2^2/g_U^{\prime 2})R^{-1}$, 
$M'_1 = (g_1^2/g_U^{\prime 2})R^{-1}$ and $M'_X = (1/2)R^{-1}$, giving 
\begin{equation}
  M'_3 : M'_2 : M'_1 : M'_X 
    = g_3^2 : g_2^2 : g_1^2 : \frac{g_U^{\prime 2}}{2},
\label{eq:KK-321-5}
\end{equation}
where $g_I$ are the standard model gauge couplings at $M_c$, and 
$g_U^{\prime 2} = g_*^2/\pi R$ with $g_*$ the 5D gauge coupling. 
The masses of the KK states for the matter and Higgs fields are 
highly model dependent, since they depend on bulk masses for 
these supermultiplets.

\subsection{Models in higher dimensions}
\label{subsec:higher-d}

In the previous subsection we have presented a model based on $SU(5)$ 
in 5D.  There are a variety of ways to extend this to higher dimensions 
and/or a larger gauge group, as was the case in higher dimensional grand 
unified theories at a high scale of order $M_U$~\cite{Kawamura:2000ev,%
Hall:2001xr}.  There are, however, new constraints in our framework.  First, 
proton decay caused by the exchange of higher dimensional unified gauge 
fields, as well as colored Higgs multiplets, must be suppressed.  This can 
be achieved, for example, by extracting different standard model multiplets 
from different bulk multiplets, as in the minimal model in the previous 
subsection.  Another constraint comes from the ($U(1)$) $R$ symmetry 
which must exist to reproduce the successful supersymmetric prediction 
for the low-energy gauge couplings.  This restricts the form of possible 
superpotential terms, giving potential constraints on the Higgs sector, 
including the sector breaking the unified symmetry (if any), as well as 
on ways of obtaining realistic fermion masses (although terms violating 
the $R$ symmetry could be generated through spontaneous breaking of $R$).

To illustrate an example of new possibilities that open up by going 
to higher dimensions, here we consider an $SU(5)$ unified theory in 6D. 
We consider that the theory possesses $N=2$ supersymmetry in 6D, which 
corresponds to $N=4$ supersymmetry in 4D, and that the extra two dimensions, 
$x^5$ and $x^6$, are compactified on a $T^2/(Z_2 \times Z'_2)$ orbifold: 
$0 \leq x^5 \leq 2\pi R_5$ and $0 \leq x^6 \leq 2\pi R_6$.  The 6D 
$N=2$ supersymmetry guarantees that the gauge anomalies in the 6D bulk 
automatically cancel.  It also requires that the only bulk field is the 
6D $SU(5)$ gauge supermultiplet, which can be decomposed into a 4D $N=1$ 
vector superfield $V$ and three 4D $N=1$ chiral superfields $\Sigma_5$, 
$\Sigma_6$ and $\Phi$, where $\Sigma_5$ and $\Sigma_6$ contain the 
fifth and sixth dimensional components of the gauge field, $A_5$ and 
$A_6$~\cite{Arkani-Hamed:2001tb}.  We now impose the following boundary 
conditions on these fields.  Along the sixth direction, $x^6$, we impose
\begin{equation}
  V(+,+), \qquad \Sigma_5(+,+), \qquad \Sigma_6(-,-), \qquad \Phi(-,-),
\label{eq:bc-x6}
\end{equation}
where $+$ and $-$ represent Neumann and Dirichlet boundary conditions, 
respectively, and the first and second signs in parentheses represent 
boundary conditions at $x^6=0$ and $x^6=\pi R_6$.  Along the fifth 
direction, $x^5$, we impose the ones in Eq.~(\ref{eq:bc-gauge-1}) 
for $V$ and $\Sigma_6$ and the ones in Eq.~(\ref{eq:bc-gauge-2}) for 
$\Sigma_5$ and $\Phi$, with the first and second signs in parentheses 
representing boundary conditions at $x^5=0$ and $x^5=\pi R_5$, 
respectively.  These boundary conditions reduce the low-energy theory 
to be the 4D $N=1$ supersymmetric $SU(3)_C \times SU(2)_L \times U(1)_Y$ 
gauge theory.  The only massless state arising from the 6D gauge 
multiplet is the 321 component of $V$.

The resulting supersymmetry and gauge symmetry structures in the extra 
two dimensions are quite rich.  There are four 5D fixed lines $x^6=0$, 
$x^6=\pi R_6$, $x^5=0$, and $x^5=\pi R_5$, each having $SU(5)$, $SU(5)$, 
321, and $SU(5)$ gauge symmetries with 5D $N=1$ (4D $N=2$) supersymmetry, 
and there are four 4D fixed points $(x^5,x^6)=(0,0), (0,\pi R_6), 
(\pi R_5,0)$, and $(\pi R_5,\pi R_6)$, each having 321, 321, $SU(5)$, 
and $SU(5)$ gauge symmetries with 4D $N=1$ supersymmetry.  The theory 
possesses a $U(1)_R$ symmetry analogous to the one in the previous 
sections, which is a linear combination of a $U(1)$ subgroup of $SU(4)_R$ 
in the 6D supersymmetry algebra and certain ``flavor'' $U(1)$'s.  The 
$U(1)_R$ charge assignment for the gauge multiplet is given by $V(0)$, 
$\Sigma_5(0)$, $\Sigma_6(0)$ and $\Phi(2)$, and that for the matter and 
Higgs fields, which are introduced on 5D or 4D subspaces, is essentially 
identical to the one in the minimal model of section~\ref{subsec:321-5}. 
Upon gauging this symmetry, we find the mixed $U(1)_R$ anomalies given 
by Eq.~(\ref{eq:U1R-anom}).  These are canceled (essentially) by a shift 
of a single modulus $M$ through the generalized Green-Schwarz mechanism, 
with the couplings of $M$ to the 321 gauge fields located at the 
$(x^5,x^6)=(0,0)$ or $(0,\pi R_6)$ fixed point (for brane-localized 
$M$).  Together with an extra moduli field $M'$ or vector-like states 
located on an $SU(5)$-preserving brane, we can have a consistent 6D 
effective field theory describing physics above the compactification 
scale $M_c$, as discussed in the previous subsection.

This setup can be used for various purposes.  Let us assume, for 
simplicity, that the shape modulus, $R_5/R_6$, is fixed (strongly) 
such that $R_5$ is (somewhat) larger than $R_6$.  In this case, the 
low-energy theory below $R_6^{-1}$ is essentially the 5D $SU(5)$ theory 
described in section~\ref{subsec:321-5}.  However, there are now a variety 
of possibilities for where to locate the fields.  For example, we can 
introduce the quark hypermultiplets $\{ {\cal Q}_i, {\cal Q}_i^c \}({\bf 
10})$, $\{ {\cal U}_i, {\cal U}_i^c \}({\bf 10})$ and $\{ {\cal D}_i, 
{\cal D}_i^c \}({\bf 5}^*)$ on the $x^6=\pi R_6$ fixed line and the 
lepton hypermultiplets $\{ {\cal L}_i, {\cal L}_i^c \}({\bf 5}^*)$ 
and $\{ {\cal E}_i, {\cal E}_i^c \}({\bf 10})$ on the $x^6=0$ fixed 
line, where $i=1,2,3$ is the generation index.  Imposing the boundary 
conditions as in Eqs.~(\ref{eq:minimal-bc-Q}~--~\ref{eq:minimal-bc-E}), 
the only low-energy states below $R_5^{-1}$ are the three generations 
of the 4D $N=1$ quark and lepton supermultiplets, $Q_i$, $U_i$, $D_i$, 
$L_i$ and $E_i$.  Locating two Higgs hypermultiplets 
$\{ {\cal H}, {\cal H}^c \}({\bf 1}, {\bf 2})_{1/2}$ and $\{ \bar{\cal H}, 
\bar{\cal H}^c \}({\bf 1}, {\bf 2})_{-1/2}$ on the $x^5=0$ fixed line, 
with the boundary conditions given by ${\cal H}(+,+)$, ${\cal H}^c(-,-)$, 
$\bar{\cal H}(+,+)$ and $\bar{\cal H}^c(-,-)$ along the sixth direction, 
we obtain two 4D $N=1$ Higgs-doublet supermultiplets $H_u$ and $H_d$ 
from these multiplets at low energies.%
\footnote{Quantization of $U(1)_Y$ hypercharges for these multiplets 
must come from physics above the cutoff scale, $M_*$.}
The Yukawa couplings $W \sim Q U H_u + Q D H_d$ and $W \sim L E H_d$ 
must be located at $(x^5,x^6)=(0,\pi R_6)$ and $(0,0)$, respectively.%
\footnote{The $S$ field can be introduced either on the $x_5=0$, $x_6=0$ 
or $x_6=\pi R_6$ fixed line, or on the $(x^5,x^6)=(0,0)$ or $(0,\pi R_6)$ 
fixed point.  The Yukawa coupling $W \sim S H_u H_d$ arises from the 
$(x^5,x^6)=(0,0)$ and/or $(0,\pi R_6)$ fixed point, while $W \sim S^3$ 
from the $(0,0)$, $(0,\pi R_6)$, $(\pi R_5,0)$ and/or $(\pi R_5,\pi R_6)$ 
fixed point.}
This setup realizes a geometrical separation between the quarks and 
leptons in an extra dimension, and thus may be used to suppresses possible 
tree-level proton decay operators.  There is, however, a possible tension 
coming from the fact that we cannot make $R_6$ very large, since it 
would increase incalculable nonuniversal contributions to the low-energy 
321 gauge couplings that arise from radiatively generated gauge kinetic 
operators on the $x^5 = 0$ line.  Thus we will still need to make some 
assumptions on the spectrum for the heavy states around the cutoff 
scale.

The story for supersymmetry breaking can be similar to that in the minimal 
model.  For $R_5$ larger than $R_6$, the $T$ field corresponds mainly 
to the modulus controlling the size for $R_5$, although it has a small 
mixture with that for $R_6$.  Assuming that the source of supersymmetry 
breaking is in nonzero $F_T$, $D_R$ and anomaly mediation, the supersymmetry 
breaking masses at the scale $M_c$ are given by Eqs.~(\ref{eq:gaugino},%
~\ref{eq:a_ABC},~\ref{eq:m2}) (although $a_{\cal H}$, $a_{\bar{\cal H}}$, 
$c_{\cal H}$ and $c_{\bar{\cal H}}$ will now be suppressed somewhat 
because the Higgs fields are brane fields in the 5D effective theory). 

Higher dimensional setups may also be used analogously to understand 
the quark and lepton masses and mixings in terms of the wavefunction 
profiles of matter fields, e.g., in the sixth dimension, although 
the issue of flavor changing neutral currents must be carefully 
addressed in such cases.  Alternatively, the 6D setup described 
here can be used to realize the scenario discussed at the end of 
section~\ref{subsec:susy-br}.  For example, we can locate all the 
quark and lepton supermultiplets on the $x^6=0$ fixed line and the 
supersymmetry breaking field $Z$ on the $(x^5,x^6)=(0,\pi R_6)$ fixed 
point.  In this case the 321 gaugino masses can be all independent, 
and the squark and slepton masses are generated through gauge 
loops and thus flavor universal.  Another use of the setup includes 
geometrically separating the $\phi$ field from the quark and lepton 
supermultiplets, ensuring the absence of potential flavor violating 
operators of the form $\int\!d^4\theta (\phi^\dagger e^{2r_\phi 
V_R} \phi) (Q_i^\dagger e^{4V_R/3} Q_j)$.

\subsection{Models with warping}
\label{subsec:warp}

So far, we have considered that the extra dimension(s) in which the 
321 gauge fields propagate is (approximately) flat.  It is, however, 
possible that there are nonnegligible warping effects.  This is, in 
fact, a natural possibility because the gauging of $U(1)_R$ is associated 
with breaking of higher dimensional Lorentz invariance.  Here we study 
the effect of warping, taking as an example the minimal 5D $SU(5)$ 
model of section~\ref{subsec:321-5}.

We consider that the 5D spacetime of the model of section~\ref{subsec:321-5} 
has a nontrivial warping.  The metric is given by
\begin{equation}
  d s^2 = e^{-2k|y|} \eta_{\mu\nu} dx^\mu dx^\nu + dy^2,
\label{eq:metric}
\end{equation}
where $k$ is the inverse curvature radius of the warped (AdS) space, 
which is taken to be somewhat smaller than the fundamental scale $M_*$. 
This corresponds to choosing the $y=0$ (321) and $y=\pi R$ ($SU(5)$) 
branes to be the ultraviolet (UV) and infrared (IR) branes, respectively. 
Unlike the case of Ref.~\cite{Randall:1999ee}, however, here we take 
the scale of the UV brane, $k$, to be of $O(10\!\sim\!100~{\rm TeV})$. 
We consider that the warp factor $e^{-\pi kR}$ is of $O(0.01\!\sim\!0.1)$, 
so that the scale of the IR brane is given by $k' \equiv k e^{-\pi kR} 
= O(1\!\sim\!10~{\rm TeV})$.

The gauge group, matter content, and boundary conditions are taken 
to be identical to those in section~\ref{subsec:321-5}.  The model then 
works analogously to the flat space case.  An important difference is 
that the wavefunction of the lightest (zero) mode of $T$ is now localized 
to the IR brane, so that its $F$-term VEV is suppressed in 4D by the 
warp factor $e^{-\pi kR}$.  This can thus be used to explain the small 
hierarchy between $M_*$ and $F_T$ ($\approx$ the weak scale) without 
using any small parameter.  The soft supersymmetry breaking masses 
are still given by Eqs.~(\ref{eq:gaugino},~\ref{eq:a_ABC},~\ref{eq:m2}), 
although the explicit expressions for $a_\Phi$ and $c_\Phi$ are changed.%
\footnote{In a warped space model, the contribution from anomaly 
mediation may be naturally suppressed $\hat{m}_C/8\pi^2 \ll \hat{m}_T, 
\hat{m}_D$.  This is because the $T$ field is expected not to feel 
(effectively) the large gravitational dimensions, as suggested by 
the ``dual'' description of the theory (given below).  The scale of 
``fundamental'' supersymmetry breaking is then small, of $O({\rm TeV})$, 
giving a very small gravitino mass, $m_{3/2} = O({\rm TeV}^2/M_{\rm Pl})$.}
Another difference is that, since the KK towers in warped space are 
localized to the IR brane, where the active gauge group is $SU(5)$, 
their spectrum is approximately $SU(5)$ symmetric.  In particular, the 
masses of the KK gauge states associated with $SU(3)_C$, $SU(2)_L$, 
$U(1)_Y$ and $SU(5)/321$ are now roughly universal $M'_3 \simeq M'_2 
\simeq M'_1 \simeq M'_X$.  Splittings among these masses, however, 
arise from the UV-brane gauge kinetic operators, which are determined 
by the observed 321 gauge couplings (for a fixed value of $\zeta'$; see 
section~\ref{subsec:321-5}).  Precise relations are given by
\begin{equation}
  M'_I \simeq M'_X 
    \left( 1 + \frac{2}{3}\frac{g_I^2}{g_U^{\prime 2} \ln(k/k')} \right),
\label{eq:KK-warped}
\end{equation}
where $M'_I$ ($I=1,2,3$) are the masses of the first KK excitations for 
the 321 gauge bosons, $M'_X$ the mass of the lightest $SU(5)/321$ gauge 
boson, $g_U^{\prime 2} = g_*^2/\pi R$ with $g_*$ the 5D $SU(5)$ gauge 
coupling, and $g_I$ the 321 gauge couplings at the scale $M'_X \simeq 
(3\pi/4)k'$.  We thus find that $M'_3 > M'_2 > M'_1 > M'_X$ with each 
mass splitting typically of about $10\%$.  (The second term in the bracket 
is $\simeq 0.25 g_I^2 (\ln(10)/\ln(k/k'))$ for $\zeta' = 1$).  The model 
described here is similar to that in Ref.~\cite{Goldberger:2002pc}, but 
now the scale of the UV brane is much smaller $k = O(10\!\sim\!100~{\rm 
TeV})$ and the gauge coupling evolution above this scale is mimicked 
by the $M$ VEV through the operators of Eq.~(\ref{eq:gen-GS-5D}).  This 
has the advantage that the 5D theory is (more) weakly coupled, since we 
can have a larger value of $M_*/\pi k$ with fixed values of the 4D gauge 
couplings.  Relatively large mass splittings of $O(10\%)$ among the 
KK gauge boson states can be a consequence of this lowered fundamental 
scale.  (The corresponding mass splittings are of order a few percent 
in the model of Ref.~\cite{Goldberger:2002pc}, since the second term 
in the bracket of Eq.~(\ref{eq:KK-warped}) is then $\approx 0.05 g_I^2$.)

The present model can be interpreted as a purely 4D theory 
(except for possible gravitational dimensions) through the AdS/CFT 
correspondence~\cite{Maldacena:1997re}.  In the 4D picture, the theory 
contains a strongly interacting (quasi-)conformal gauge sector $G$, 
whose conformality is spontaneously broken at the scale $k'$.  The 
$G$ sector possesses an $SU(5)$ global ``flavor'' symmetry, of which 
the $SU(3) \times SU(2) \times U(1)$ subgroup is explicitly gauged 
and identified as the standard model gauge group 321.  There are 
quark, lepton and Higgs supermultiplets, $Q_i$, $U_i$, $D_i$, $L_i$, 
$E_i$, $H_u$, $H_d$ and $S$, which are charged under 321 (except 
for $S$) and neutral under $G$.  At the fundamental scale $M_*$, the 
gauge kinetic terms (gauge couplings) of 321 come purely from the 
terms of the form ${\cal L} = -\sum_{I=1,2,3} \int\!d^2\theta\, 
(\zeta_I \langle M \rangle/c)\, {\cal W}_I^{a\alpha} {\cal W}_{I\alpha}^a 
+ {\rm h.c.}$, where $\zeta_I = A_I - \zeta'$ with $\zeta' \geq 1$. 
The remaining universal piece then comes from an asymptotically 
non-free contribution from the $G$ sector: $\delta(1/g_I^2)|_G = 
(b_G/8\pi^2) \ln(k/k')$, where $b_G$ ($ > 0$) is the beta-function 
coefficient for the contribution from $G$, and the universality of the 
contribution is guaranteed by the global $SU(5)$ symmetry.  The required 
numerology can be read off from Eqs.~(\ref{eq:req-M},~\ref{eq:req-T}) 
in the case of $\zeta'=1$: $\langle M \rangle/c \simeq 0.25$ and 
$(b_G/8\pi^2)\ln(k/k') \simeq 1$.  Note that the understanding of 
the coefficient $\zeta_I$ is quite simple in this context.  The 
quantities $A_I$ and $-\zeta'$ represent the mixed $U(1)_R$-321 
gauge anomalies carried by the elementary (quark, lepton, Higgs 
and 321 gauge) states and the $G$ states, respectively, and the sum 
of them, $A_I-\zeta' = \zeta_I$, is canceled by the shift of $M$. 
Supersymmetry breaking is caused by the IR dynamics of the $G$ 
sector at the scale $k'$, and is transmitted to the MSSM (and $S$) 
states through mixings between the elementary and $G$ states and 
by 321 gauge loops.

The 4D description of the theory given above allows us to make a simple 
estimate for the scales appearing in the theory.  First, from the 
observed values of the 321 gauge couplings, we find that the contribution 
from the $G$ sector to the 321 gauge couplings should be
\begin{equation}
  \delta\frac{1}{g_I^2}\Biggr|_G = 
    \frac{b_G}{8\pi^2}\ln\frac{k}{k'} \simeq 2-\zeta'.
\label{eq:G-contr}
\end{equation}
Let us now focus on the case with $\zeta' = 1$ for simplicity, although 
similar results are also obtained for other values of $\zeta'$ unless 
$\zeta'$ is tuned such that $\delta(1/g_I^2)|_G \ll 1$.  (Note that 
$\zeta' \simlt 2$ to reproduce the observed 321 gauge couplings with 
$b_G > 0$).  Since $k/k' = e^{\pi kR} = O(10\!\sim\!100)$ in the present 
setup, Eq.~(\ref{eq:G-contr}) gives $b_G \approx (20\!\sim\!30)$, implying 
that the $G$ sector is quite ``large'' when viewed from the 321 sector. 
(This is important to guarantee that the 5D picture is weakly coupled.) 
Now, let us assume that supersymmetry breaking is induced by the strong 
$G$ dynamics that do not involve any particularly small or large numbers. 
In this case, the mass of the superparticles can be estimated using 
large-$N$ scaling~\cite{'tHooft:1973jz}; in particular, we find that 
the 321 gaugino masses $M_I$ are given by
\begin{equation}
  M_I \simeq \frac{g_I^2}{16\pi^2} b_G M_\rho,
\label{eq:MI-est}
\end{equation}
where $M_\rho \approx \pi k'$ represents the mass scale for the resonances 
in the $G$ sector, i.e. the mass scale for the KK towers in the 5D picture. 
(For similar analyses, see~\cite{Nomura:2004zs}.)  This equation implies 
that for a large $G$ sector of $b_G \approx (20\!\sim\!30)$, the mass 
hierarchy between the superparticles and the KK resonances is only of 
a factor of $\approx 10$.  We thus expect that the masses of the KK gauge 
bosons, which are approximately $SU(5)$ symmetric, are (only) of order 
a few TeV.  Note that in the conventional desert case, e.g. in the model 
of~\cite{Goldberger:2002pc}, the large desert of $\ln(k/k') \simeq 30$ 
leads to $b_G \simlt (4\!\sim\!5)$, making the KK masses much higher. 
This opens the exciting possibility that  some of these states may be 
observed at the LHC.  Moreover, the value of $b_G$ may be measured from 
mass splittings among these states.  Using the AdS/CFT correspondence, 
we find that Eq.~(\ref{eq:KK-warped}) can be written as $M'_I \simeq 
M'_X (1 + g_I^2 b_G/12\pi^2)$.  If we find a value of $b_G$ that is 
significantly larger than $\simeq 5$ from these mass relations (or 
through Eq.~(\ref{eq:MI-est})), it would provide a strong experimental 
suggestion, possibly together with large values of $\lambda$ and/or 
$\kappa$, that the fundamental scale of nature is not much far above 
the weak scale.

\section{Discussion and Conclusions}
\label{sec:concl}

Unraveling the physical origin of electroweak symmetry breaking will 
be one of the central themes in particle physics in the next few years. 
The LHC will start exploring an energy region well above the masses 
of the electroweak gauge bosons within two years, which may reveal 
some new physics beyond the standard model whose existence is intimately 
related to electroweak symmetry breaking.  It is then crucial that we 
extract as much information as possible from this data both experimentally 
and theoretically.  In particular, it will be important to explore 
possible interpretations of the data, which may eventually allow us 
to uncover some basic aspects of the fundamental theory of nature.

Supersymmetry is one of the leading candidates for new physics at 
the electroweak scale.  Its successes are often stated in the context 
of the supersymmetric desert --- weak scale supersymmetry stabilizes 
the large hierarchy between the Planck and the weak scales, and it leads 
to a successful unification of the three gauge couplings at a high energy 
close to the Planck scale.  In our view, the most {\it robust} success 
of supersymmetry, however, lies in the fact that it stabilizes the 
hierarchy between the weak scale and the scale that suppresses most 
of higher dimension operators of the standard model.  From the LEP data, 
we know that this (small) hierarchy almost certainly exists.  Supersymmetry 
provides one of the best ways to stabilize it without leading to an 
immediate conflict with the precision electroweak data, since the 
electroweak symmetry is still broken by the VEVs of Higgs fields that 
are perturbative at the weak scale.  While aspects associated with the 
existence of the desert may well be an illusion, being the result of 
a vast extrapolation in many orders of magnitude in energy, we feel 
that the feature of supersymmetry just described should play a role 
if weak scale supersymmetry is actually realized in nature.

In this paper we have studied a framework of weak scale supersymmetry 
in which (most of) the virtues of the supersymmetric desert are naturally 
reproduced without actually having a large energy interval above the 
weak scale.  Such a picture may, in fact, be suggested from naturalness 
of electroweak symmetry breaking, since fine-tuning in conventional 
supersymmetric theories often arises from the large logarithmic running 
of supersymmetry breaking masses and/or the conflict of a large coupling(s) 
with the Landau pole constraint, neither of which applies in the absence 
of the large desert.  We have shown that a (gauged) $U(1)_R$ symmetry 
that assigns the same charge for all the matter and Higgs supermultiplets 
may play an important role --- it may reproduce the successful supersymmetric 
prediction for the low-energy gauge couplings because of its relation to 
conformal symmetry.  We have demonstrated that this can indeed be realized 
in effective field theory, and have constructed classes of explicit 
models based on higher dimensional unified field theories.  The $U(1)_R$ 
symmetry can have important consequences for the form of the observable 
sector superpotential; in particular, the Higgs sector superpotential 
is expected not to contain any dimensionful parameters.  This allows, 
together with a low fundamental scale, for making the mass of the 
lightest Higgs boson rather large, $\approx (200\!\sim\!300)~{\rm GeV}$, 
helping to eliminate fine-tuning in electroweak symmetry breaking.  The 
consistency with the precision electroweak data can be recovered through 
the contributions from the Higgs sector and/or new states at a multi-TeV 
region, such as the KK states associated with the standard model gauge 
fields.

There are many natural sources of supersymmetry breaking in this framework 
-- the auxiliary component VEVs of various singlet (moduli) fields, $U(1)_R$ 
$D$-term VEV, and anomaly mediation.  An interesting aspect here is that 
what is usually referred to (broadly) as gravity mediation now occurs 
at a scale not much far above the weak scale.  We have calculated the 
resulting pattern of supersymmetry breaking masses, and find that it can 
be quite distinct.  For example, in the case that these masses are dominated 
by the $U(1)_R$ $D$-term VEV and the auxiliary component VEV of the field 
giving the universal contribution to the standard model gauge couplings, 
we obtain gaugino masses proportional to the square of the gauge couplings 
and squark and slepton masses proportional to their anomalous dimensions 
at the scale where these masses are generated (between of order a few 
and a hundred TeV).  Since such a pattern of superparticle masses does 
not arise very naturally in the conventional desert framework, its 
observation at the LHC may provide a nontrivial hint for the absence 
of the desert.

Finally we comment on possible variations of the basic framework presented 
in this paper.  We have mainly considered the fundamental scale, $M_*$, 
to be in the range of $O(10\!\sim\!100~{\rm TeV})$.  It should, however, 
be obvious that there is no real upper bound on this scale, except that 
it should probably be smaller than the 4D (not reduced) Planck scale 
$\approx 10^{19}~{\rm GeV}$.  While lower values of $M_*$ may be preferred 
from naturalness of electroweak symmetry breaking, there is nothing 
really wrong with other values; for example, the case with intermediate 
scale $M_*$ may be an interesting possibility.%
\footnote{An alternative, amusing application of our models is to 
consider $M_*$ to be at the gravitational (or the conventional string) 
scale with $R^{-1}$ close to $M_*$.  The effective gauge coupling 
unification scale can then be {\it lower} than $M_*$ if $\langle M 
\rangle/c$ is negative, reproducing the conventional unification scale 
of $M_U \simeq 2 \times 10^{16}~{\rm GeV}$ for $\langle M \rangle/c 
\approx -(0.03\!\sim\!0.05)$.  Such a value of $\langle M \rangle/c$ 
can, in fact, be generated through stabilization discussed in this 
paper or through standard gaugino condensation.}
Such a variation of $M_*$ will have interesting consequences on the 
superparticle spectrum and the resulting phenomenology.  The charge 
assignment for $U(1)_R$ may also be modified.  For example, one may 
consider a family-dependent charge assignment, which does not modify 
the prediction for the gauge couplings as long as it commutes with 
$SU(5)$.  The Yukawa couplings are then generated through the VEV of 
the $\phi$ field, which may partly explain the origin of the observed 
Yukawa hierarchy if the $\phi$ VEV is somewhat smaller than the cutoff 
scale, due to, e.g., a large $U(1)_R$ charge for $\phi$.  Theories 
which use a non-$R$ $U(1)$ gauge symmetry, instead of a $U(1)_R$ 
gauge symmetry, can also be considered, and we present this possibility 
in Appendix~\ref{app:non-R}.  Finally, the ``large'' gravitational 
dimension may be strongly warped, in which case our basic picture 
may have to be modified.  For example, we can attach a strongly warped 
gravitational dimension to the basic module of section~\ref{sec:framework}. 
The resulting theory can then be a 5D supersymmetric theory with the 
metric given by Eq.~(\ref{eq:metric}) in which all the MSSM and $S$ 
states are localized to the IR brane at $y=\pi R$.  The scales at 
the UV and IR branes can be chosen to be of order the 4D Planck scale 
and $(10\!\sim\!100)~{\rm TeV}$, respectively.  This picture is then 
close to that of Ref.~\cite{Randall:1999ee}, but with the small hierarchy 
between the IR-brane cutoff scale and the weak scale stabilized by 
supersymmetry and with the successful gauge coupling prediction reproduced 
through interactions of the form of Eq.~(\ref{eq:gen-GS}).  The way 
this theory avoids the Landau pole constraints is similar to that 
in~\cite{Birkedal:2004zx}, although the absence of the standard model 
gauge fields in the bulk allows for the 5D theory to be more weakly 
coupled.  Note that this theory in fact possesses a large energy desert, 
as can be clearly seen in the 4D dual picture.  The desert, however, 
is simply not relevant for the MSSM and $S$ states in this picture 
because they are composite states generated at the scale of 
$(10\!\sim\!100)~{\rm TeV}$.

In summary, we have presented a simple and realistic framework for 
supersymmetry which does not possess a large energy desert above the 
weak scale.  The framework has rich phenomenological implications, and 
allows for detailed analyses of, e.g., electroweak symmetry breaking. 
If the LHC finds superparticles whose spectrum shows features discussed 
in this paper and/or if it discovers a Higgs sector which indicates 
some of the couplings are so large that they hit the Landau pole before 
the conventional unification scale, then it would provide a strong 
suggestion that the fundamental scale of nature may not be far above 
the weak scale.  It is exciting that we may indeed be able to explore 
this possibility in the next few years.

\section*{Acknowledgments}

This work was supported in part by the Director, Office of Science, 
Office of High Energy and Nuclear Physics, of the US Department of 
Energy under Contract DE-AC03-76SF00098, by the National Science 
Foundation under grant PHY-0403380, by a DOE Outstanding Junior 
Investigator award, and by an Alfred P. Sloan Research Fellowship.

\appendix

\section{Soft Supersymmetry Breaking Parameters in the Minimal Model}
\label{app:soft}

In this appendix we present explicit expressions for the soft 
supersymmetry breaking terms in the model of section~\ref{subsec:321-5}. 
The gaugino masses are given by Eq.~(\ref{eq:gaugino}).  The scalar 
trilinear interactions are given in general by Eq.~(\ref{eq:a_ABC}). 
Defining ${\cal L}_{\rm soft} = -(a_u)_{ij} \tilde{q}_i \tilde{u}_j 
h_u - (a_d)_{ij} \tilde{q}_i \tilde{d}_j h_d - (a_e)_{ij} \tilde{l}_i 
\tilde{e}_j h_d - (a_\nu)_{ij} \tilde{l}_i \tilde{n}_j h_u - a_\lambda 
s h_u h_d - (a_\kappa/3) s^3 + {\rm h.c.}$, this gives
\begin{eqnarray}
  (a_u)_{ij} &=& -(y_u)_{ij} 
    \Bigl( a_{\cal H} \hat{m}_T 
      + (\gamma_{Q_i}+\gamma_{U_j}+\gamma_{H_u}) \hat{m}_C \Bigr),
\label{eq:a_u} \\
  (a_d)_{ij} &=& -(y_d)_{ij} 
    \Bigl( a_{\bar{\cal H}} \hat{m}_T 
      + (\gamma_{Q_i}+\gamma_{D_j}+\gamma_{H_d}) \hat{m}_C \Bigr),
\label{eq:a_d} \\
  (a_e)_{ij} &=& -(y_e)_{ij} 
    \Bigl( a_{\bar{\cal H}} \hat{m}_T 
      + (\gamma_{L_i}+\gamma_{E_j}+\gamma_{H_d}) \hat{m}_C \Bigr),
\label{eq:a_e} \\
  \Bigl[ (a_\nu)_{ij} &=& -(y_\nu)_{ij} 
    \Bigl( a_{\cal H} \hat{m}_T 
      + (\gamma_{L_i}+\gamma_{N_j}+\gamma_{H_u}) \hat{m}_C \Bigr) \Bigr],
\label{eq:a_nu} \\
  a_\lambda &=& -\lambda 
    \Bigl( (a_{\cal S} + a_{\cal H} + a_{\bar{\cal H}}) \hat{m}_T
      + (\gamma_{S}+\gamma_{H_u}+\gamma_{H_d}) \hat{m}_C \Bigr),
\label{eq:a_lambda} \\
  a_\kappa &=& -3 \kappa (a_{\cal S} \hat{m}_T + \gamma_S \hat{m}_C),
\label{eq:a_kappa}
\end{eqnarray}
where $a_\varphi$ ($\varphi = {\cal H}, \bar{\cal H}, {\cal S}$) are given by
$a_\varphi = 2\pi R M_\varphi/(e^{2\pi R M_\varphi}-1)$ (see Eq.~(\ref{eq:a_phi})). 
Here, we have used Eq.~(\ref{eq:localization}), and the neutrino Yukawa couplings 
$(y_\nu)_{ij}$ in Eq.~(\ref{eq:a_nu}) are defined analogously to the other 
Yukawa couplings in Eq.~(\ref{eq:Yukawa}).

The soft supersymmetry breaking scalar squared masses are given by 
Eq.~(\ref{eq:m2}).  For the squarks and sleptons, this gives
\begin{eqnarray}
  m_{\tilde{F}_i}^2 &=& -\gamma_{F_i} \hat{m}_D^2 
    + \frac{1}{2} \frac{d\gamma_{F_i}}{d\ln\mu_R} |\hat{m}_C|^2
    - \sum_I \frac{C_I^F}{8\pi^2} \frac{g_I^4}{g_U^{\prime 2}}
      \Bigl( \hat{m}_T \hat{m}_C^* + {\rm h.c.} \Bigr),
\label{eq:m2_Fi} \\
  m_{\tilde{Q}_3}^2 &=& -\gamma_{Q_3} \hat{m}_D^2 
    + \frac{1}{2} \frac{d\gamma_{Q_3}}{d\ln\mu_R} |\hat{m}_C|^2
    + \frac{1}{8\pi^2} \Biggl( \frac{1}{2} (y_t^2 a_{\cal H} + y_b^2 
        a_{\bar{\cal H}}) - \sum_I C_I^Q \frac{g_I^4}{g_U^{\prime 2}} \Biggr)
      \Bigl( \hat{m}_T \hat{m}_C^* + {\rm h.c.} \Bigr),
\label{eq:m2_Q3} \\
  m_{\tilde{U}_3}^2 &=& -\gamma_{U_3} \hat{m}_D^2 
    + \frac{1}{2} \frac{d\gamma_{U_3}}{d\ln\mu_R} |\hat{m}_C|^2
    + \frac{1}{8\pi^2} \Biggl( y_t^2 a_{\cal H}
      - \sum_I C_I^U \frac{g_I^4}{g_U^{\prime 2}} \Biggr)
      \Bigl( \hat{m}_T \hat{m}_C^* + {\rm h.c.} \Bigr),
\label{eq:m2_U3} \\
  m_{\tilde{D}_3}^2 &=& -\gamma_{D_3} \hat{m}_D^2 
    + \frac{1}{2} \frac{d\gamma_{D_3}}{d\ln\mu_R} |\hat{m}_C|^2
    + \frac{1}{8\pi^2} \Biggl( y_b^2 a_{\bar{\cal H}}
      - \sum_I C_I^D \frac{g_I^4}{g_U^{\prime 2}} \Biggr)
      \Bigl( \hat{m}_T \hat{m}_C^* + {\rm h.c.} \Bigr),
\label{eq:m2_D3} \\
  m_{\tilde{L}_3}^2 &=& -\gamma_{L_3} \hat{m}_D^2 
    + \frac{1}{2} \frac{d\gamma_{L_3}}{d\ln\mu_R} |\hat{m}_C|^2
    + \frac{1}{8\pi^2} \Biggl( \frac{1}{2} (y_\tau^2 a_{\bar{\cal H}} 
        + y_{\nu_3}^2 a_{\cal H}) - \sum_I C_I^L \frac{g_I^4}{g_U^{\prime 2}} 
      \Biggr) \Bigl( \hat{m}_T \hat{m}_C^* + {\rm h.c.} \Bigr),
\label{eq:m2_L3} \\
  m_{\tilde{E}_3}^2 &=& -\gamma_{E_3} \hat{m}_D^2 
    + \frac{1}{2} \frac{d\gamma_{E_3}}{d\ln\mu_R} |\hat{m}_C|^2
    + \frac{1}{8\pi^2} \Biggl( y_\tau^2 a_{\bar{\cal H}} 
      - \sum_I C_I^E \frac{g_I^4}{g_U^{\prime 2}} \Biggr)
      \Bigl( \hat{m}_T \hat{m}_C^* + {\rm h.c.} \Bigr),
\label{eq:m2_E3} \\
  \Biggl[ m_{\tilde{N}_3}^2 &=& -\gamma_{N_3} \hat{m}_D^2 
    + \frac{1}{2} \frac{d\gamma_{N_3}}{d\ln\mu_R} |\hat{m}_C|^2
    + \frac{1}{8\pi^2} y_{\nu_3}^2 a_{\cal H} 
      \Bigl( \hat{m}_T \hat{m}_C^* + {\rm h.c.} \Bigr) \Biggr],
\label{eq:m2_N3}
\end{eqnarray}
where $F=Q,U,D,L,E$ (and $N$), the index $i=1,2$ runs over the first two 
generations in Eq.~(\ref{eq:m2_Fi}), and $C_I^F$ are the group theoretical 
factors given by $(C_1^F, C_2^F, C_3^F) = (1/60,3/4,4/3)$, $(4/15,0,4/3)$, 
$(1/15,0,4/3)$, $(3/20,3/4,0)$, $(3/5,0,0)$ and $(0,0,0)$ for $F = Q, U, D, L, 
E$ and $N$, respectively.    Here, we have retained only the third generation 
Yukawa couplings $y_t \equiv (y_u)_{33}$, $y_b \equiv (y_d)_{33}$, $y_\tau 
\equiv (y_e)_{33}$ and $y_{\nu_3} \equiv (y_\nu)_{33}$ (which may not 
be valid for the neutrino Yukawa couplings).  We have also used one-loop 
expressions for $\gamma_\Phi$ in the last terms in the right-hand-sides 
of Eqs.~(\ref{eq:m2_Fi}~--~\ref{eq:m2_N3}) for illustrative purposes.
The first and second terms are the contributions from a nonvanishing 
$U(1)_R$ $D$-term VEV and pure anomaly mediation, respectively.

For the Higgs fields $h_u$, $h_d$ and $s$, we find
\begin{eqnarray}
  m_{h_u}^2 &=& -\gamma_{H_u} \hat{m}_D^2 + c_{\cal H} |\hat{m}_T|^2
    + \frac{1}{2} \frac{d\gamma_{H_u}}{d\ln\mu_R} |\hat{m}_C|^2
\nonumber\\
  && + \frac{1}{8\pi^2} \Biggl( \frac{1}{2} \Bigl\{ (3 y_t^2 + y_{\nu_3}^2 
        + \lambda^2) a_{\cal H} + \lambda^2 a_{\bar{\cal H}} + \lambda^2 
        a_{\cal S} \Bigr\} - \sum_I C_I^H \frac{g_I^4}{g_U^{\prime 2}} \Biggr)
      \Bigl( \hat{m}_T \hat{m}_C^* + {\rm h.c.} \Bigr),
\label{eq:m2_Hu} \\
  m_{h_d}^2 &=& -\gamma_{H_d} \hat{m}_D^2 + c_{\bar{\cal H}} |\hat{m}_T|^2
    + \frac{1}{2} \frac{d\gamma_{H_d}}{d\ln\mu_R} |\hat{m}_C|^2
\nonumber\\
  && + \frac{1}{8\pi^2} \Biggl( \frac{1}{2} \Bigl\{ (3 y_b^2 + y_\tau^2 
        + \lambda^2) a_{\bar{\cal H}} + \lambda^2 a_{\cal H} + \lambda^2 
        a_{\cal S} \Bigr\} - \sum_I C_I^H \frac{g_I^4}{g_U^{\prime 2}} \Biggr)
      \Bigl( \hat{m}_T \hat{m}_C^* + {\rm h.c.} \Bigr),
\label{eq:m2_Hd} \\
  m_{s}^2 &=& -\gamma_{S} \hat{m}_D^2 + c_{\cal S} |\hat{m}_T|^2
    + \frac{1}{2} \frac{d\gamma_{S}}{d\ln\mu_R} |\hat{m}_C|^2
\nonumber\\
  && + \frac{1}{8\pi^2} \Bigl\{ (\lambda^2 + 3\kappa^2) a_{\cal S} 
        + \lambda^2 a_{\cal H} + \lambda^2 a_{\bar{\cal H}} \Bigr\}
      \Bigl( \hat{m}_T \hat{m}_C^* + {\rm h.c.} \Bigr),
\label{eq:m2_S}
\end{eqnarray}
where $(C_1^H, C_2^H, C_3^H) = (3/20,3/4,0)$.  The coefficients $c_\varphi$ 
($\varphi = {\cal H}, \bar{\cal H}, {\cal S}$) are given (in flat space) 
by $c_\varphi = (\pi R M_\varphi/\sinh(\pi R M_\varphi))^2$ (see 
Eq.~(\ref{eq:c_phi})).

\section{Theories with a Non-{\boldmath $R$} {\boldmath $U(1)$} Gauge Symmetry}
\label{app:non-R}

In this appendix we present an alternative class of theories in which 
a non-$R$ pseudo-anomalous $U(1)$ gauge symmetry is used instead of the 
gauged $R$ symmetry.  This class of theories has a potential advantage 
in that the cutoff scale does not have to be 
``charged'' under the $U(1)$ gauge symmetry, i.e. one can regulate the 
theory in such a way that there is no ``anomalous'' transformation of 
the superspace density under a supersymmetric $U(1)$ transformation. 
This may allow us to rely less on the structure of the ultraviolet theory 
above $M_*$ to explain the form of the low-energy effective Lagrangian.%
\footnote{For example, it may be easier in these theories to conceive 
that $U(1)$ invariance of the (observable sector) superspace density 
is ensured purely by the $U(1)$ gauge multiplet $V$, while that 
of the gauge kinetic functions (at the quantum level) by a moduli 
field $M$.  This will be the case, e.g., if the matter and Higgs 
fields are geometrically separated from the $M$ field in extra spatial 
dimensions in which the standard model gauge fields propagate.}
As we will see below, this is achieved at the expense of somewhat 
arbitrary choices of the matter content and the $U(1)$ charge assignment.

Let us first consider the effective theory below $M_c$.  We introduce, 
as usual, the standard model quark, lepton, and Higgs chiral superfields, 
$Q_i$, $U_i$, $D_i$, $L_i$, $E_i$, $H_u$ and $H_d$ ($i=1,2,3$), which 
have the Yukawa couplings of Eq.~(\ref{eq:Yukawa}).  We then introduce 
a (pseudo-anomalous) $U(1)_A$ gauge symmetry, under which the matter 
and Higgs fields carry charges of $+1/2$ and $-1$, respectively.  While 
the Yukawa couplings are invariant under this charge assignment, the 
``fundamental'' mass term for the Higgsinos, $W = \mu H_u H_d$, is not. 
We thus introduce a singlet field $S(+2)$ together with the interaction 
$W = \lambda S H_u H_d$, which gives a necessary mass term for the 
Higgsinos after $S$ obtains a VEV and also provides an extra contribution 
to the physical Higgs boson mass. 

At this point, the mixed $U(1)_A$-321 gauge anomalies are given by 
$A_1 = 12/5$, $A_2 = 2$ and $A_3 = 3$, which do not take the necessary 
form of $A_I \propto b_I + b$ ($I=1,2,3$), where $b_I$ are the MSSM 
beta function coefficients, $(b_1, b_2, b_3) = (33/5, 1, -3)$, and 
$b$ is a constant which does not depend on the gauge group $I$.  To 
fix this ``problem,'' we introduce fields $X_1$, $X_2$ and $X_3$ which 
transform as adjoints under the 321 gauge group, i.e. $X_1({\bf 1}, 
{\bf 1})_0$, $X_2({\bf 1}, {\bf 3})_0$ and $X_3({\bf 8}, {\bf 1})_0$, 
and carry a $U(1)_A$ charge of $+3$.  The mixed anomalies are then given 
by $A_1 = 12/5$, $A_2 = 8$ and $A_3 = 12$, which satisfies
\begin{equation}
  A_I = -b_I + 9.
\label{eq:AI-bI-rel}
\end{equation}
These mixed anomalies are canceled by terms of the form of 
Eq.~(\ref{eq:gen-GS}) upon introduction of a moduli field $M$ that 
transforms as $M \rightarrow M + (c/16\pi^2)\Lambda$ under $U(1)_A$, 
where $\Lambda$ is the gauge transformation parameter superfield for 
$U(1)_A$ in the normalization that a chiral superfield with a $U(1)_A$ 
charge $q$ transforms as $\Phi \rightarrow e^{q \Lambda} \Phi$.  (The 
$U(1)_Y$-$U(1)_A^2$ anomaly is automatically vanishing, and we do not 
consider the $U(1)_A^3$ or $U(1)_A$-$({\rm gravity})^2$ anomalies since 
they depend on unknown fields that are singlet under the 321 gauge 
group.)  Introducing the universal contribution $\langle T \rangle$ 
to the 321 gauge kinetic functions, as in Eq.~(\ref{eq:gauge-kin}), 
we find that the 321 gauge couplings at $M_c$, $g_I$, take exactly 
the same form as that arising in the conventional supersymmetric 
desert picture, Eq.~(\ref{eq:321-GUT}).  The correspondence between 
the two theories is now given by
\begin{eqnarray}
  \langle T \rangle - \frac{36}{c} \langle M \rangle 
    &\leftrightarrow& \frac{1}{g_U^2},
\label{eq:nonR-corresp-1}\\
  \langle M \rangle 
    &\leftrightarrow& \frac{c}{32\pi^2}\ln\frac{M_U}{M_c},
\label{eq:nonR-corresp-2}
\end{eqnarray}
instead of Eqs.~(\ref{eq:corresp-1},~\ref{eq:corresp-2}), and the relation 
among the low-energy gauge couplings is given by the standard supersymmetric 
one in Eq.~(\ref{eq:gcu}).  The required VEVs of $\langle T \rangle = O(1) > 0$ 
and $\langle M \rangle/c = O(0.1) > 0$ can be generated in the way discussed 
in section~\ref{subsec:M-stab}.  With $c = O(1)$, this can be done within 
the regime of effective field theory.

In order for the model to be realistic, the adjoint fields $X_1$, $X_2$ and 
$X_3$ must obtain masses of order the weak scale or larger (at least for $X_2$ 
and $X_3$).  We thus introduce a singlet field $\phi$ with a $U(1)_A$ charge 
of $-6$, together with the superpotential interactions $W = (\lambda_1/2) 
\phi X_1^2 + (\lambda_2/2) \phi X_2^2 + (\lambda_3/2) \phi X_3^2$, where 
the couplings $\lambda_I$'s may or may not respect the $SU(5)$ relation, 
$\lambda_1 = \lambda_2 = \lambda_3$, at a scale $\approx M_c$.  The $\phi$ 
field obtains a VEV of order $M_*/4\pi$ to cancel the Fayet-Iliopoulos term 
for $U(1)_A$.  The Fayet-Iliopoulos term is generated after the $M$ 
field obtains a VEV, since its K\"ahler potential is given by $K = M_*^2\, 
{\cal F}(M + M^\dagger + (c/8\pi^2) V_A)$, where $V_A$ is the $U(1)_A$ 
gauge multiplet and ${\cal F}(x)$ is an arbitrary polynomial in $x$ 
with coefficients of $O(1)$ up to symmetry factors.  We assume that the 
generated Fayet-Iliopoulos term is positive, in which case $\phi$ can 
absorb (most of) this term.  The fact that $\phi$ absorbs, and not 
$H_u$ and $H_d$, should be determined energetically after supersymmetry 
breaking effects are included.  We can alternatively generate a larger 
VEV for $\phi$, of order $M_*$, if we introduce the superpotential 
$W = Y(\phi\bar{\phi} - M_*^2)$, where $Y$ and $\bar{\phi}$ are chiral 
superfields with $U(1)_A$ charges of $0$ and $+6$, respectively. 
A nonvanishing $\langle \phi \rangle$ gives masses to the adjoint 
fields $X_1$, $X_2$ and $X_3$.  It also provides the $S^3$ term 
in the superpotential through $W = \eta \phi S^3/M_*$. 

Summarizing, the superpotential of the model is given by
\begin{equation}
  W = W_{\rm Yukawa} + \lambda S H_u H_d + \frac{\eta}{3M_*} \phi S^3 
    + \frac{\lambda_1}{2} \phi X_1^2 + \frac{\lambda_2}{2} \phi X_2^2 
    + \frac{\lambda_3}{2} \phi X_3^2,
\label{eq:nonR-W}
\end{equation}
where $\phi$ obtains a nonvanishing VEV of $O(M_*/4\pi\!\sim\!M_*)$ and 
$W_{\rm Yukawa}$ is given by Eq.~(\ref{eq:Yukawa}).  The $U(1)_A$ charge 
assignment for the fields is given by
\begin{equation}
\begin{array}{cccccccccccccc}
  & Q_i & U_i & D_i & L_i & E_i & (N_i) & H_u & H_d & S & \phi & X_1 & X_2 & X_3 \\
  U(1)_A: & 1/2 & 1/2 & 1/2 & 1/2 & 1/2 & 1/2 & -1 & -1 & 2 & -6 & 3 & 3 & 3 
\end{array}.
\label{eq:nonR-charge}
\end{equation}
An interesting point is that, despite the absence of $U(1)_R$, we are again 
led to the form of the superpotential of the next-to-minimal supersymmetric 
standard model.  (The absence of $M$ in the superpotential is assumed, as 
always.)  Some of the dangerous higher dimension operators in the superpotential, 
such as the one leading to large Majorana neutrino masses, are suppressed 
by the (broken) $U(1)_A$ symmetry.  The standard matter parity ($R$ parity) 
also arises automatically as a discrete subgroup of $U(1)_A$, if the the 
K\"ahler potential is made $U(1)_A$ invariant purely by the $U(1)_A$ gauge 
multiplet, $V_A$, and not by the combination $-(8\pi^2/c)(M+M^\dagger)$. 
(The stability of protons can be ensured in a way discussed in 
section~\ref{subsec:others}.)  Note that a constant term in the 
superpotential, necessary to cancel the cosmological constant, can 
be introduced without breaking $U(1)_A$. 

The model can be embedded into higher dimensions analogously to 
section~\ref{sec:model}.  Both flat space and warped space models 
can be constructed.  The $U(1)_A$ charge assignment for 5D supermultiplets 
$\{ \Phi, \Phi^c \}$ ($\Phi = Q_i, U_i, \cdots$) is determined by 
Eq.~(\ref{eq:nonR-charge}), with a conjugated field carrying the 
opposite $U(1)_A$ charge from a non-conjugated field.  With the help 
of an extra moduli field $M'$ or vector-like states located on the 
$SU(5)$-preserving brane, we can have a consistent effective field 
theory describing physics above the compactification scale $M_c$. 

The story for supersymmetry breaking is similar to the $U(1)_R$ 
case.  We have in general contributions from $F_T$, $F_M$, anomaly 
mediation, as well as a nonvanishing $D$-term for $U(1)_A$, $D_A$, 
which is generated, e.g., if the $\phi$ field has a supersymmetry 
breaking mass of order the weak scale.  In the case that the contributions 
arise from $F_T$, $D_A$ and anomaly mediation, the supersymmetry 
breaking masses at the scale $M_c$ are given by Eqs.~(\ref{eq:gaugino},%
~\ref{eq:a_ABC},~\ref{eq:m2}) but with the first term in the right-hand-side 
of Eq.~(\ref{eq:m2}) replaced as
\begin{equation}
  - \gamma_\Phi \hat{m}_D^2 \rightarrow 
    q_\Phi \hat{m}_D^2,
\label{eq:nonR-D}
\end{equation}
where $q_\Phi$ is the $U(1)_A$ charge of a superfield $\Phi$, 
and $\hat{m}_D^2$ is now given by $\hat{m}_D^2 = -D_A$ instead 
of Eq.~(\ref{eq:mD2}).  In the presence of gauge kinetic mixing 
between $U(1)_A$ and $U(1)_Y$ at tree level, this term is given 
by $(q_\Phi - \epsilon g_1^2 Y_\Phi) \hat{m}_D^2$, where $\epsilon$ 
is the coefficient of the kinetic mixing term and $Y_\Phi$ the 
$U(1)_Y$ hypercharge of the chiral superfield $\Phi$ in the 
$SU(5)$ normalization (see footnote~\ref{ft:kin-mix}).

\end{document}